\pdfoutput=1

\documentclass[11pt,twoside,a4paper,cmspaper,final,collab]{cms-tdr}
\def\svnVersion{315673}\def\cmsCernNo{2013-037}\def\cmsCernDate{\today}\def\cmsMessage{Published in Physics Letters B as \href{http://dx.doi.org/10.1016/j.physletb.2015.09.062}{\doi{10.1016/j.physletb.2015.09.062.}}}

\begin{document}\cmsNoteHeader{HIG-14-006}

\hyphenation{had-ron-i-za-tion}
\hyphenation{cal-or-i-me-ter}
\hyphenation{de-vices}
\RCS$Revision: 235553 $
\RCS$HeadURL: svn+ssh://svn.cern.ch/reps/tdr2/notes/HIG-14-006/trunk/HIG-14-006.tex $
\RCS$Id: HIG-14-006.tex 235553 2014-04-07 06:37:43Z soffi $

\newlength\cmsFigWidth
\ifthenelse{\boolean{cms@external}}{\setlength\cmsFigWidth{0.48\textwidth}}{\setlength\cmsFigWidth{0.75\textwidth}}
\ifthenelse{\boolean{cms@external}}{\providecommand{\cmsLeft}{top}\xspace}{\providecommand{\cmsLeft}{left}\xspace}
\ifthenelse{\boolean{cms@external}}{\providecommand{\cmsRight}{bottom}\xspace}{\providecommand{\cmsRight}{right}\xspace}
\cmsNoteHeader{HIG-14-006}
\title{Search for diphoton resonances in the mass range from 150 to \texorpdfstring{850\GeV}{850 GeV} in pp collisions at \texorpdfstring{$\sqrt{s} = 8\TeV$}{sqrt(s)=8 TeV}}

\date{\today}

\newcommand {\Hgg}{\ensuremath{\PH \to \gamma \gamma}\xspace}
\newcommand {\Zee}{\ensuremath{\Z \to \Pep\Pem}\xspace}
\newcommand {\mgg}{\ensuremath{m_{\gamma \gamma}}\xspace}
\newcommand{\PX}{\ensuremath{\cmsSymbolFace{X}}\xspace}
\newcommand{\cosba}{\ensuremath{\cos(\beta -\alpha)}\xspace}
\newcommand{\PA}{\ensuremath{\cmsSymbolFace{A}}\xspace}
\newcommand{\PV}{\ensuremath{\cmsSymbolFace{V}}\xspace}
\providecommand{\CL}{CL\xspace}
\providecommand{\CLs}{\ensuremath{\mathrm{CL}_\mathrm{s}}\xspace}

\abstract{
Results are presented of a search for heavy particles decaying into two photons.
The analysis is based on a 19.7\fbinv sample of proton-proton collisions at $\sqrt{s}=8\TeV$ collected with the CMS detector at the CERN LHC.
The diphoton mass spectrum from 150 to 850\GeV is used to search for an excess of events over the background.
The search is extended to new resonances with natural widths of up to 10\% of the mass value.
No evidence for new particle production is observed and limits at 95\% confidence level
on the production cross section times branching fraction to diphotons are determined. These limits are interpreted in terms of two-Higgs-doublet model parameters.
}

\hypersetup{%
pdfauthor={CMS Collaboration},%
pdftitle={Search for diphoton resonances in the mass range from 150 to 850 GeV in pp collisions at sqrt(s) = 8 TeV},%
pdfsubject={CMS},%
pdfkeywords={Higgs, BSM, diphoton, new scalar, 2HDM, graviton, Randall--Sundrum, ADD models},
}

\maketitle

\section{Introduction}\label{Sec:intro}
The discovery of a standard model-like Higgs boson at the CERN LHC~\cite{Chatrchyan:2013lba,Khachatryan:2014ira,higgsdiscoveryAtlas,Aad:2014eha}
opens a new phase in the understanding of the
standard model (SM) of particle physics.
The search for additional Higgs-like particles and the measurement of their properties
 provide complementary ways to test the validity of the SM and to test for the presence of physics beyond it.

This analysis describes a search for new resonances in the diphoton invariant mass spectrum,
using data corresponding to an integrated luminosity of 19.7\fbinv collected with the CMS detector at a center-of-mass energy of 8\TeV at the CERN LHC.
Despite the large nonresonant background, the diphoton decay mode
provides a clean final-state topology that allows the mass of the
decaying object to be reconstructed with high precision, exploiting the excellent performance of the
electromagnetic calorimeter of the CMS experiment.
The analysis searches for local excesses that could be due to
the production of  particles that decay into two photons with mass in the range
from 150 to 850\GeV.
Both narrow and wide resonances are investigated with natural widths ranging from 100 \MeV to 10\% of the resonance mass.
This search covers the diphoton mass range above that investigated in \cite{Khachatryan:2014ira,Aad:2014eha}.
The ATLAS experiment recently published a similar search for narrow resonances in the diphoton final state in the mass
range between 65 and 600\GeV at a center-of-mass energy of 8\TeV~\cite{PhysRevLett.113.171801}.
Previous  searches for resonant diphoton processes have been performed at the Tevatron by D0~\cite{Abazov:2010xh} and CDF~\cite{Aaltonen:2010cf} at a center-of-mass energy of 1.96\TeV and
by the ATLAS~\cite{Aad:2012cy} and CMS~\cite{Chatrchyan:2011fq} experiments at the LHC at a center-of-mass energy of 7\TeV.

Several models of physics beyond the SM, such as the two-Higgs-doublet model (2HDM)~\cite{Craig:2013hca},
motivate the search for additional high-mass resonances in the diphoton channel.
Generally, these models provide an extension of the Higgs sector, where a total of five Higgs bosons
are predicted by the theory.
The mass spectrum of the 2HDM can be split into two regions: a light SM-like Higgs boson \Ph with mass around 125\GeV
and the remaining physical Higgs bosons, $\PH$, a scalar, $\PA$, a pseudoscalar, and $\Hpm$, clustered at an equal or higher scale with $m_\PH \sim m_\PA \sim m_{\Hpm}$.
Under the assumption that the newly observed Higgs boson is the light CP-even Higgs scalar of the 2HDM,
the consistency of its couplings with those predicted by the SM pushes the model close to the so called alignment limit \cite{Dev:2014yca}, where
certain decay modes of heavy
neutral Higgs bosons vanish, including $\PH \to \PV \PV$ (where V is a vector boson), $\PH \to \Ph\Ph$, and $\PA \to Z \Ph$.
At the same time, decays of \PH and \PA to \Pgg \Pgg\ and \Pgt \Pgt\ become increasingly important and the electroweak production modes,
such as vector boson fusion or production in association with a \PW\ or a $\Z$ boson, are predicted to be  suppressed.
Therefore the production of both \PH and \PA is dominated by gluon fusion.
The absence of tree-level flavor-changing neutral currents in multiple-Higgs-doublet theories is guaranteed by the Glashow--Weinberg condition~\cite{PhysRevD.15.1958}.
This condition is satisfied in the 2HDM by four discrete combinations of the Yukawa couplings of the fermions to the Higgs doublets.
In the Type~I scenario all fermions couple to one doublet, while in Type~II up-type quarks couple to one doublet and down-type quarks and leptons couple to the other.
A detailed description of other scenarios is given in Ref.~\cite{Craig:2013hca}.

Given the general character of this search, the results can also be interpreted in terms of different spin hypotheses for the new particle.
The Landau--Yang theorem~\cite{Landau,PhysRev.77.242} forbids the direct decay of a spin-1 particle into a pair of photons.
The cases of spin-0 and spin-2 are investigated in this analysis.
Spin-2 particles decaying into two photons are predicted by other extensions of the SM
such as the Randall--Sundrum~\cite{graviton} and the Arkani-Hamed--Dimopoulos--Dvali~\cite{ArkaniHamed:1998rs} models.
These theories predict a distinct higher-dimensional scenario,
 which provides an approach to the hierarchy problem alternative to supersymmetry.
The particle predicted in this context, the graviton, can have a mass in the \TeVns range and thus
be observed at the LHC preferentially in its decay into two gauge bosons, such as photons.

The paper is organized as follows: after a brief description of the CMS detector in Section~\ref{sec:cms},
the data and simulated samples are described in Section~\ref{sec:samples},
while the reconstruction and identification of photons is detailed in Section~\ref{sec:phot}.
 The diphoton vertex identification is covered in Section~\ref{sec:vertex}, followed by the description
of the event selection and classification in Section~\ref{sec:evclasses}.
 Sections~\ref{sec:sig} and~\ref{sec:bkg} describe, respectively, the signal and background models used
 for the interpretation of the data, and Section~\ref{sec:syst} discusses the associated systematic uncertainties.
Finally, in Section~\ref{sec:res} the model independent results of the search for new diphoton resonances and their interpretation in terms of
the standalone production and decay rates for \PH and \PA within the 2HDM are discussed.
We express these results in terms of the appropriate 2HDM parameters.
\section{The CMS detector}\label{sec:cms}
The central feature of the CMS detector is a superconducting solenoid, of 6 m internal diameter,
 providing an axial magnetic field of 3.8\unit{T} along the beam direction.
 Within the field volume there are a silicon pixel and strip tracker, a lead tungstate crystal electromagnetic calorimeter (ECAL),
 and a brass and scintillator hadron calorimeter (HCAL). Charged particle trajectories are measured by the tracker system, covering $0 \le \phi \le 2\Pgp$ in azimuth and $\abs{\eta}<2.5$ in pseudorapidity.

 Muons are measured in gas-ionization detectors embedded in the steel return yoke.
The ECAL, which surrounds the tracker volume, consists of 75\,848 lead tungstate crystals that provide coverage in pseudorapidity $\abs{\eta} < 1.48$ in the barrel region (EB)
and $1.48 < \abs{\eta} < 3.00$ in two endcap regions (EE). The EB modules are arranged in projective
towers. A preshower detector consisting of two planes of silicon sensors interleaved with a
total of 3 radiation lengths of lead is located in front of the EE. In the region $\abs{\eta} < 1.74$, the HCAL cells have
widths of 0.087 in pseudorapidity and azimuth. In the  $( \eta, \phi )$  plane, and for $\abs{\eta} < 1.48$, the
HCAL cells map onto $5 \times 5$ ECAL crystal arrays to form calorimeter towers projecting radially
outward from close to the nominal interaction point. At larger values of $\abs{\eta}$, the size of the
towers increases and the matching ECAL arrays contain fewer crystals. Within each tower, the
energy deposits in ECAL and HCAL cells are summed to define the calorimeter tower energies,
subsequently used to calculate the energies and directions of hadronic jets.
A more detailed description of the CMS detector, together with a definition of the coordinate system used and the relevant kinematic variables,
 can be found in Ref.~\cite{Chatrchyan:2008zzk}.

Photons generally deposit their energy in a group of crystals of the ECAL called  ``cluster".
Reconstruction of the photons used in this analysis is described in Section~\ref{sec:phot}, and uses a clustering
of the energy recorded in the ECAL, known as a ``supercluster", which may be extended in
the $\phi$ direction to form an extended cluster or group of clusters~\cite{paperphot}.
Several procedures are used to calibrate
the energy response of individual crystals before the clustering steps \cite{Chatrchyan:2013dga}.
The changes in the transparency of the ECAL crystals due to irradiation during the LHC running
periods and their subsequent recovery are monitored continuously, and corrected for, using light injected from a laser system.
The calibration of the ECAL is achieved by exploiting the $\phi$ symmetry of the energy flow, and by using
photons from $\Pgpz \to \Pgg \Pgg $ and $\Pgh \to \Pgg \Pgg $  decays
and electrons from $\PW \to \Pe \Pgne$ and $\Z \to \EE$ decays~\cite{Chatrchyan:2013dga}.
\section{Data samples and simulated events}\label{sec:samples}
The events used in the analysis are selected by two diphoton triggers
with asymmetric transverse momentum thresholds ($\pt$), 26 and 18\GeV or 36 and 22\GeV
on the leading and subleading photons respectively, depending on the data taking period, and complementary photon selections.
One selection requires a loose calorimetric identification based on the shape of
the electromagnetic shower and loose isolation requirements on the photon candidates,
while the other requires only that the photon candidates have a high value of the $\RNINE$ shower shape variable.
The $\RNINE$ variable is defined as the energy sum of $3\times 3$ crystals centered on the most energetic crystal in
the supercluster divided by the energy of the supercluster.
Photons that convert before reaching the calorimeter
tend to have wider showers and lower values of $\RNINE$ than unconverted photons.
High trigger efficiency is maintained by allowing both photons to satisfy either selection.
The measured trigger efficiency is above 99.8\% for events satisfying the diphoton preselection discussed in Section~\ref{sec:presel}
required for events entering the analysis.

Monte Carlo (MC) signal and background events are generated using a combination of programs.
Full detector response is simulated with \GEANTfour~\cite{GEANT}.
Multiple simultaneous pp interactions (pileup) are simulated,
and the events are weighted to reproduce the distribution of primary vertices observed in data.
The interactions used to simulate pileup are generated with the same
version of \PYTHIA~\cite{PYTHIA}, v6.424, that is used for other purposes as described below.
The \PYTHIA tune used for the underlying event activity is Z2*~\cite{Chatrchyan:2011id}.
New resonances $\PX$ are simulated with a natural width of 0.1\GeV which is smaller than the value of the mass resolution in the energy range considered.
Both spin-0 and spin-2 hypotheses are considered.
Simulated signal samples of $\PX \to \Pgg \Pgg$ events are generated with  \PYTHIA~\cite{PYTHIA}
for the gluon fusion process with the following mass hypotheses: 150, 200, 250, 300, 400, 600 and 850\GeV.
Interference between the signal and the background is studied using \SHERPA~2.1.0~\cite{Gleisberg:2008ta,Dixon:2013haa}.
The exclusion limits detailed in Section~\ref{sec:res}  are computed using either a signal plus interference
template given by \SHERPA or the Breit--Wigner theoretical model described in Section~\ref{sec:sig}.
The inclusion of the interference in our signal model affects the expected and observed upper limits on the signal yields at the level of 3\% or less.
The effect is therefore negligible compared to the systematic uncertainties, described in Section~\ref{sec:syst}, and is not included in the final results.
Simulated backgrounds include the diphoton continuum process involving two prompt photons,
generated with  \SHERPA~1.4.2~\cite{Gleisberg:2008ta}, and processes where one of the photon candidates arises from
misidentified jet fragments, simulated with \PYTHIA. These two contributions represent 97\% of the total background in this analysis.
A less than 3\% contribution is expected from QCD events in which both photons candidates arise from misidentified jet fragments; this contribution is neglected in the analysis.

\section{Photon reconstruction and identification}\label{sec:phot}
Photon candidates for the analysis are reconstructed from energy deposits in the ECAL using
algorithms that constrain the superclusters in $\eta$ and $\phi$ to the shapes expected from photons with high $\pt$.
The clustering algorithms account for about 98\% of the energy of the photons,
including those that undergo conversion and bremsstrahlung in the material in front of the ECAL.
Groups of clusters are used to form superclusters.
In the barrel region, superclusters are formed from five-crystal-wide strips in $\eta$,
centered on the locally most energetic crystal (seed), and have a variable extension in $\phi$
to take into account the effect of the magnetic field on electrons from photons showering before ECAL.
In the endcaps, where the crystals are arranged according to an $x$-$y$ rather than an $\eta$-$\phi$ geometry,
matrices of $5\times 5$ crystals, which may partially overlap and are centered on the seed crystal,
are summed if they lie within a narrow $\phi$ road.
About half of the photons convert into \EE pairs in the material upstream of the ECAL.
If the resulting charged-particle tracks originate sufficiently close to the interaction point to pass
through three or more tracking layers, conversion track pairs may be reconstructed and matched to the photon candidate.

The energy containment of the photon showers in the clustered crystals and the shower losses due
to conversions in the material upstream of the calorimeter are corrected using a multivariate
regression technique based on a boosted decision tree (BDT)~\cite{2005NIMPA.543..577R,2007physics...3039H},
which uses as input a collection of shower shape and kinematic variables, together with the energy measured in the preshower for events with photons in the endcaps.
Corrections are derived from simulation.
In order to correct for residual discrepancies between simulation and data, the MC simulation
is tuned to match the energy resolution observed in data, while data are calibrated to match the energy response in simulation.
More details about photon reconstruction can be found in~\cite{paperphot}.

The photon candidates used in the analysis are required to be within the fiducial region, $\abs{\eta} < 2.5$,
excluding the barrel-endcap transition region $1.44 < \abs{\eta} < 1.57$.
\subsection{Photon preselection}\label{sec:presel}
The photon candidates in this analysis are required to satisfy preselection criteria similar to
the trigger requirements:
\begin{itemize}
\item $\pt^{\Pgg_1} >33\GeV$ and $\pt^{\Pgg_2} > 25$\GeV, where $\pt^{\Pgg_1}$ and $\pt^{\Pgg_2}$ are the transverse momenta of
the leading and subleading photons, respectively;
\item a selection on the hadronic leakage of the shower, measured as the ratio of energy in
HCAL cells behind the supercluster to the energy in the supercluster;
\item a photon identification based on isolation and shape of the shower with lower thresholds with respect  the one used in the final photon selection described in Section \ref{sec:phid};
\item an electron veto, which removes the photon candidate if its supercluster is
 matched to an electron track with no missing hits in the innermost tracker layers.
\end{itemize}
The efficiency of the photon preselection, measured in data using a ``tag-and-probe'' technique~\cite{CMS:2011aa}, ranges from 94\% to 99\%~\cite{Khachatryan:2014ira}.
The efficiency of all preselection criteria, except the electron veto requirement, is measured using  $\Z\to\EE$ events, while
the efficiency for photons to satisfy the electron veto requirement
is measured using $\Z\to \mu\mu\gamma$ events, in which the photon is produced by final state
radiation, which provide a more than 99\% pure source of prompt photons.
\subsection{Photon identification and selection}\label{sec:phid}
Photon identification is performed by applying selection requirements on a set of discriminating
variables.
In this analysis, the selection is optimized separately in four different pseudorapidity and $\RNINE$ regions ~\cite{Khachatryan:2014ira}.
The variables used to suppress the background due to the misidentification of jets with high electromagnetic content are
\begin{itemize}
\item the sum of the transverse momenta of all the tracks coming from the vertex chosen for the event (described
in Section \ref{sec:vertex}) within a veto cone of
 $\DR = \sqrt{\smash[b]{(\Delta\eta)^2+(\Delta\phi)^2}} = 0.3$
around the photon candidate's direction;
\item the sum of the ECAL energy deposits in crystals located within a veto cone of $\DR = 0.3$ around the supercluster position, excluding the photon;
\item the sum of the energies of HCAL towers whose centers lie within an annular veto region
of outer radius $\DR_o=0.4$ and inner radius $\DR_i= 0.15$, centered on the ECAL supercluster position;
\item the ratio between the sum of HCAL tower energies within a veto cone of size $\DR<0.15$ centered
on the ECAL supercluster position, and the energy of the supercluster;
\item the spread in $\eta$ of the electromagnetic cluster,  computed with logarithmic weights and defined as
\begin{equation}\begin{gathered}
\sigma^2_{\eta \eta} = \frac{\sum w_i(\eta_i-\bar{\eta})^2}{\sum w_i}, \text{ where} \\
\bar{\eta} = \frac{\sum w_i\eta_i}{\sum w_i} \text{ and } w_i = \max\left(0, 4.7+\ln \frac{E_i}{E_{5{\times}5}}\right),
  \end{gathered}\end{equation}
and the sum runs over the $5{\times}5$  crystal matrix around the most energetic crystal in the supercluster. $E_{5{\times}5}$ is the energy of the $5{\times}5$  crystal matrix and $\eta_i = 0.0174 \hat{\eta}_i$, where $\hat{\eta}_i$ is the $\eta$ index of the $i$th crystal.
This variable represents the second moment of the energy distribution along the $\eta$ coordinate.
\end{itemize}
A more detailed description of the isolation requirements can be found in~\cite{paperphot}.
The energy deposited within the isolation cones is corrected for the contribution from pileup and the underlying event
using the \FASTJET technique~\cite{Cacciari2008119}.
\subsection{Diphoton vertex identification}\label{sec:vertex}
The diphoton mass resolution has contributions from the resolution of
the measurement of the photon energies and the angle
between the two photons.
The opening angle resolution strongly depends on
the determination of the interaction point where the two photons were produced.
If the vertex from which the photons originate
is known with a precision better than 10\mm, the experimental resolution on the
angle between them makes a negligible contribution to the mass resolution.

No charged particle tracks result from photons that do not convert, so the diphoton vertex is identified indirectly,
using the kinematic properties of the diphoton system and its correlations with the kinematic properties of the
recoiling tracks. If either of the photons converts, the direction of the resulting tracks can provide additional information~\cite{Khachatryan:2014ira}.

The efficiency for finding the correct vertex for a diphoton resonance of mass above 150\GeV is between 79\% and 92\% and increases with the mass of the resonance.
\section{Event selection and classification}\label{sec:evclasses}
The analysis uses events with two photon candidates satisfying the
preselection and identification requirements, and
with $\pt^{\Pgg_1} > m_{\Pgg\Pgg}/3$  and $\pt^{\Pgg_2} > m_{\Pgg\Pgg}/4$.
The use of $\pt$ thresholds scaled by $m_{\Pgg\Pgg}$ prevents
a distortion of the low end of the $m_{\Pgg\Pgg}$ spectrum that would result from fixed thresholds~\cite{Chatrchyan:2012twa}.
This strategy has two other main advantages.
Background modeling is simplified because the shapes of the invariant mass
distributions in the different event classes are similar after this selection.
In addition scaled $\pt$ thresholds allow for tighter selection criteria with
the mass increase, while preserving the acceptance of the selection.

The search sensitivity is increased by subdividing the events into classes,
according to indicators of mass resolution and predicted signal-to-background ratio.
Two simple classifiers are used: the minimum $\RNINE$
and the maximum pseudorapidity of the two photons.
Photons with a high value of the $\RNINE$ variable are predominantly unconverted, have a better
energy resolution than those with a lower value, and are
less likely to arise from misidentification of jet fragments. Similarly, photons in the
barrel have better energy resolution than those in the endcap and are less likely to be incorrectly identified.
The classification scheme groups together events with good diphoton mass resolution, resulting from photons
with good energy resolution and with better signal-to-background ratio.
The event class definitions are shown in Table~\ref{tab:CiC-classes}.
\begin{table}[tb]
  \topcaption{Definition of diphoton event classes}
  \label{tab:CiC-classes}
  \centering
  \begin{tabular}{cccc}
    Class & $\eta$  criterion & $\RNINE$ criterion \\
    \hline
   0               &$\max(\abs{\eta})<1.44$ & $\min(\RNINE)>0.94$ \\
    1               & $\max(\abs{\eta})<1.44$ & $\min(\RNINE)<0.94$ \\
   2               &$1.57 < \max(\abs{\eta})<2.50 $& $\min(\RNINE)>0.94$ \\
   3               & $1.57  < \max( \abs{\eta} )<2.50$& $\min(\RNINE)<0.94$ \\
  \end{tabular}
\end{table}

\begin{figure}[tb]
 \begin{center}
\includegraphics[width=\cmsFigWidth]{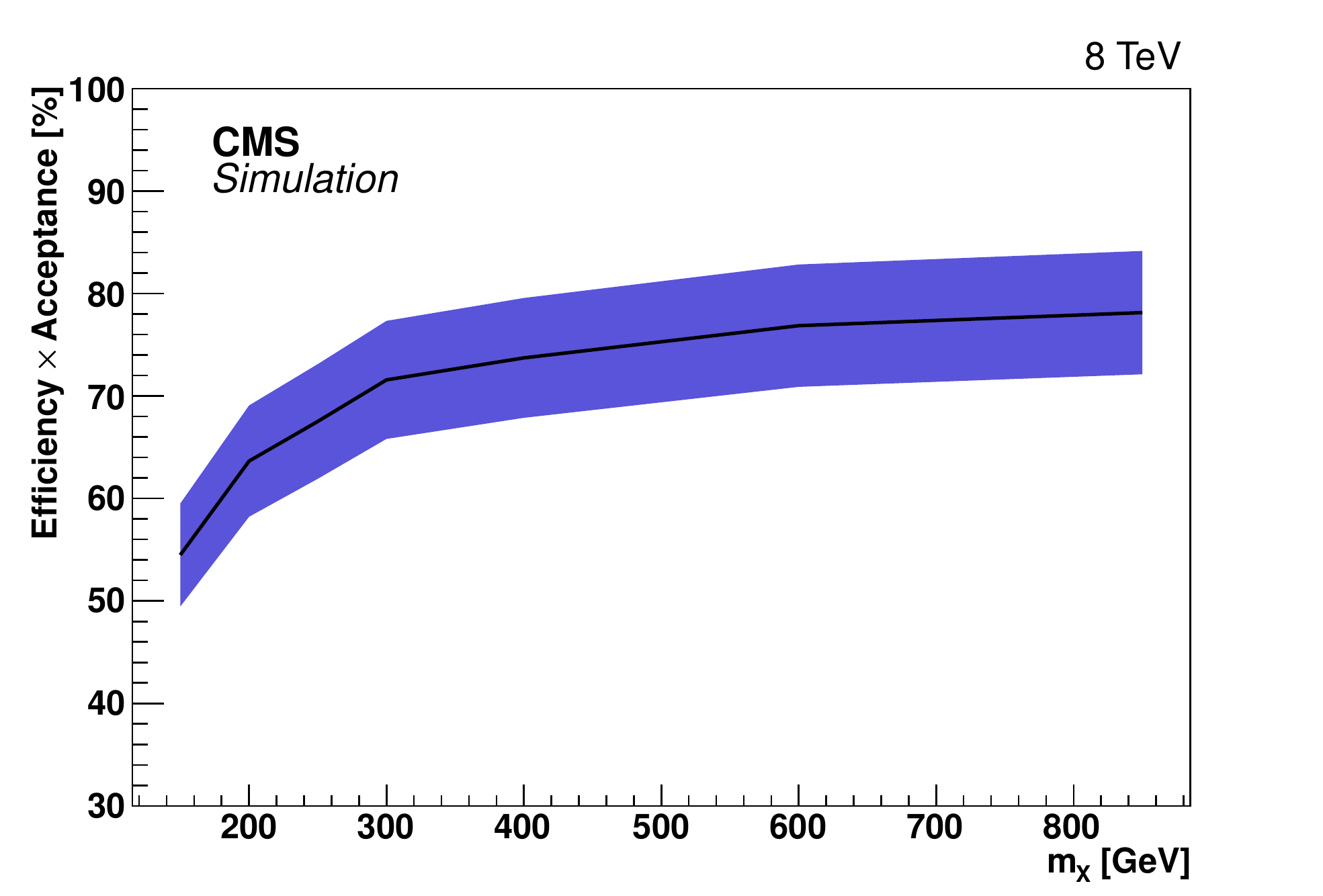}
  \caption{ Signal efficiency times acceptance as a function of the mass for a spin-0 scalar resonance, produced via gluon-gluon fusion, with natural width equal to 0.1\GeV. The shaded region indicates the systematic uncertainty.}
  \label{fig:effacc}
 \end{center}
\end{figure}

\begin{figure}[tbh]
 \begin{center}
\includegraphics[width=\cmsFigWidth]{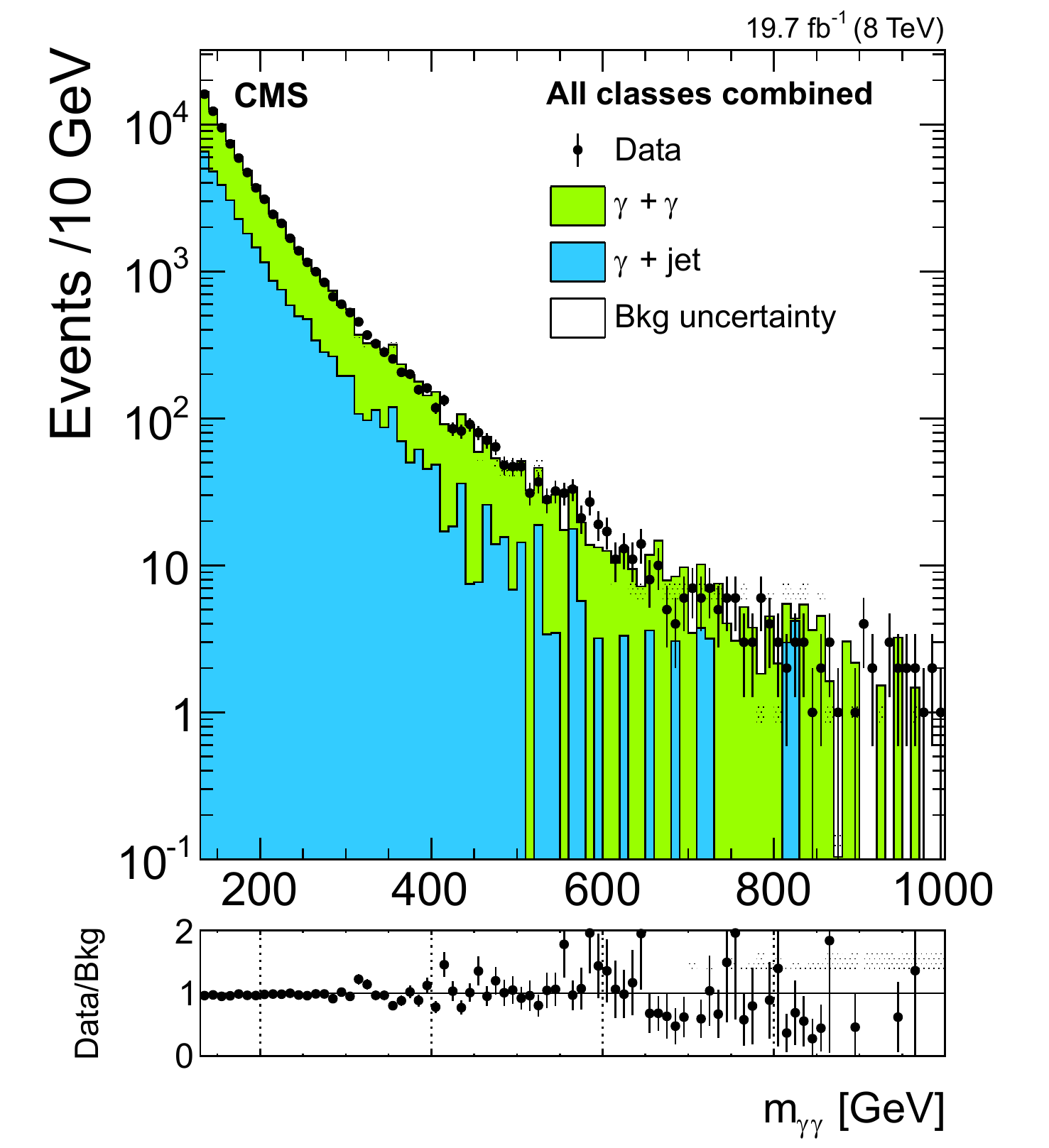}
  \caption{Diphoton invariant mass distribution for the selected events in data and simulation.
  Background processes are represented by the filled histograms. The shaded band represents the Poisson uncertainty in the MC prediction.The prompt diphoton events and the photon plus
jets events are shown separately. The ratio between data and MC is displayed bin by bin in the bottom plot.  In this figure, the leading-order cross sections for the background processes are scaled to next-to-leading-order (NLO) predictions with scale factors derived from CMS measurements at 7\TeV~\cite{Chatrchyan:2011qt,Chatrchyan:2011qta}.
}
  \label{fig:invMassComb}
 \end{center}
\end{figure}

Figure~\ref{fig:effacc} shows the signal efficiency times the acceptance of the event selection as a function of
the mass hypothesis for a narrow spin-0 scalar resonance produced via gluon-gluon fusion.
The shaded region shows the systematic uncertainty due to different sources as described in Section \ref{sec:syst}.
The acceptance over the full mass range is also evaluated for the pseudoscalar hypothesis and found to be compatible with
that measured for the scalar resonances within the uncertainties.
For the spin-2 scenario the corresponding
numbers are about 5--10\% smaller, because of differences in acceptance between the two models.
Figure~\ref{fig:invMassComb} shows the diphoton invariant mass distribution for selected events in data and MC simulation, normalized to an integrated luminosity of 19.7\fbinv, for all event classes combined.
The bin width of this distribution  is chosen to be narrow enough to properly display wide resonances.
However a data driven technique is exploited in this analysis for the estimation of the background,  as detailed in Section~\ref{sec:bkg}.
\section{Statistical methodology}\label{sec:method}
In order to assess the compatibility of the data with the presence of a diphoton resonance, we make a hypothesis test based on a frequentist construction~\cite{Read:2000ru,Junk:1999kv}.
An unbinned maximum-likelihood fit  in a sliding window range to the diphoton
invariant mass distributions in all the event classes is made using a
parametric model for the signal and a background shape obtained directly from data.
A scan of the signal mass and the signal width is performed.
The signal and background normalizations are allowed to float, together with the parameters which describe the background shape. The log-likelihood ratio is used as test statistic.
The signal model is derived from MC simulation as described in Section~\ref{sec:sig},
with corrections determined from data-MC comparisons applied.
The functional form of the background distribution is determined by fitting the measured $m_{\Pgg \Pgg}$ distribution as detailed in Section~\ref{sec:bkg}.
Systematic uncertainties described in Section~\ref{sec:syst}
are incorporated into the analysis via nuisance parameters and are treated according to the frequentist paradigm~\cite{CMS-NOTE-2011-005}.

\subsection{Signal parametrization}\label{sec:sig}
To construct the signal fitting function from simulated events, the reconstructed $m_{\Pgg \Pgg}$ distribution
of each event class is described in terms of a parametric function of
the hypothetical signal mass $m_{\PX}$.
This procedure can be extended in a simple fashion to allow for an additional free parameter,
the natural width $\Gamma_{\PX}$ of the new resonance,
by convolving a resolution function, which takes into account
the detector response, with a theoretical line shape.
The convolution of the detector response, described in this section, with the theoretical lineshape,
described in section~\ref{sec:resp}, is applied for any natural width hypothesis.

\begin{figure}[tbh]
 \begin{center}
\includegraphics[width=0.48\textwidth]{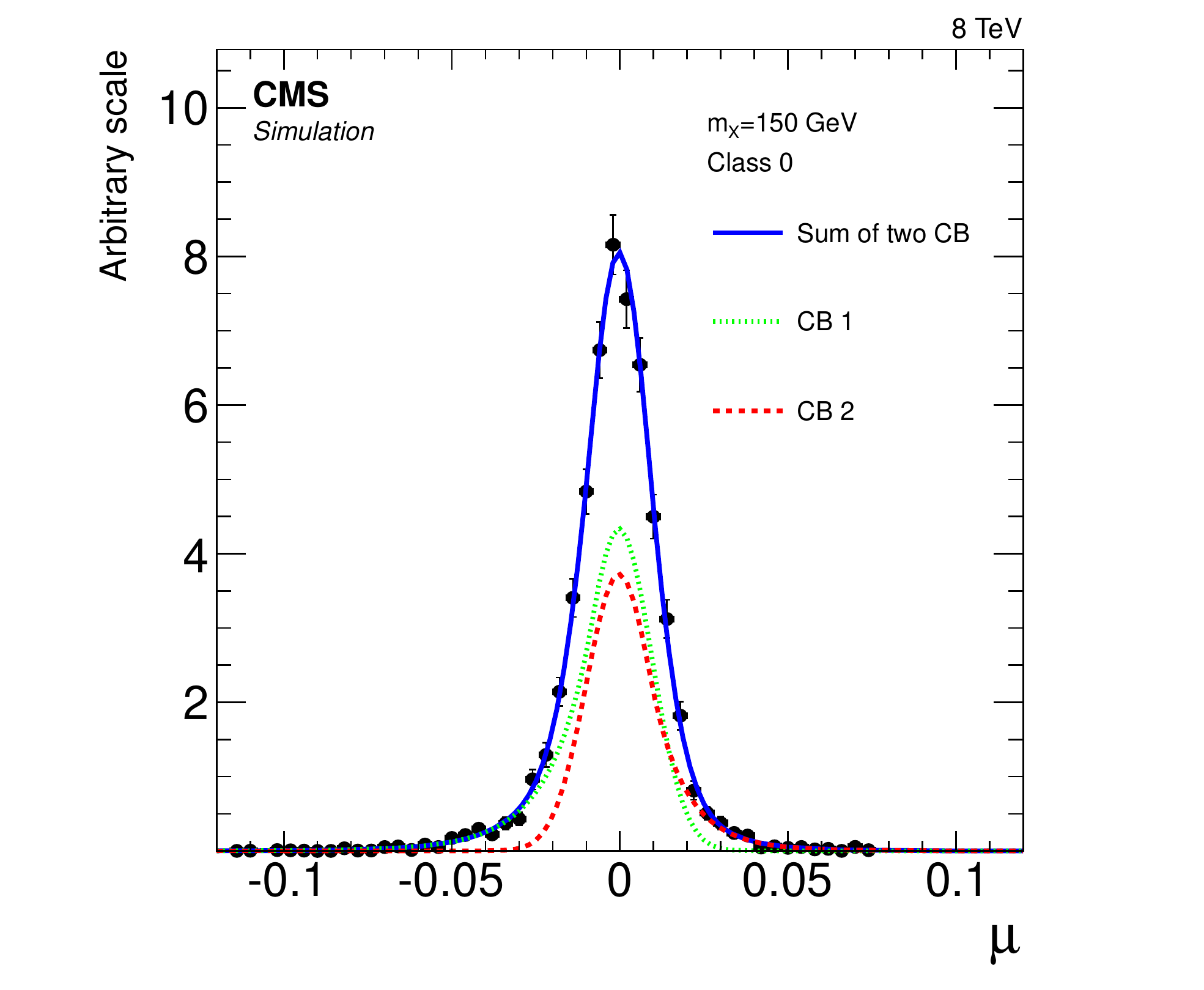}
\includegraphics[width=0.48\textwidth]{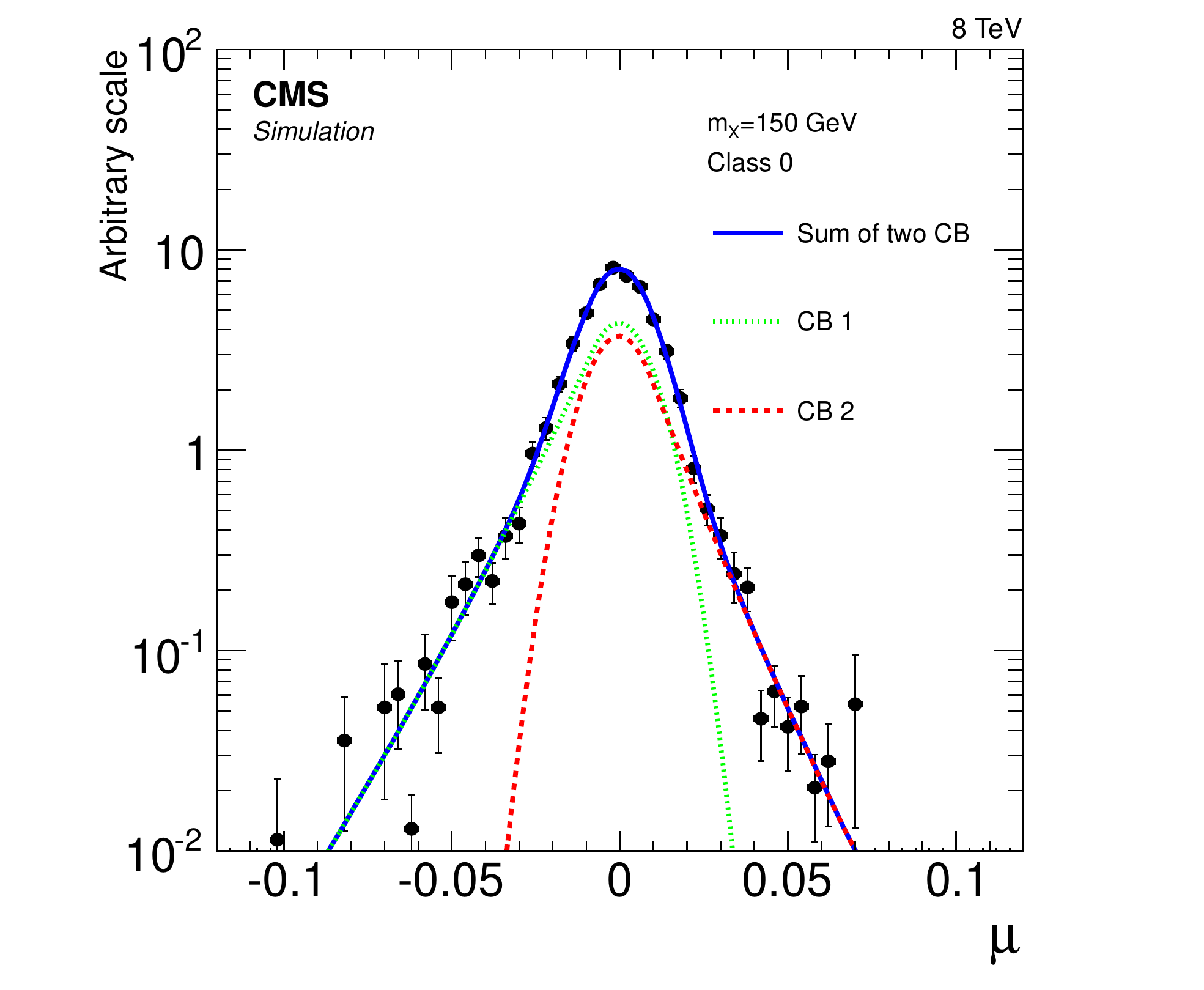}
  \caption{Fit to the detector response distribution for $m_\PX = 150\GeV$ with two single-sided Crystal Ball functions (solid curve), displayed with linear (\cmsLeft) and logarithmic (\cmsRight) scales. The dashed and dotted curves show the individual Crystal Ball components.}
 \label{fig:fitRes}
 \end{center}
\end{figure}

The detector response is parametrized in terms of the relative difference between the reconstructed diphoton mass $m_\text{reco}$ and the true mass $m_\text{true}$,  $\mu = (m_\text{reco}-m_\text{true})/m_\text{true}$.
The resolution function $R$ is obtained by fitting the response distribution with an analytic function,
namely the sum of two single-sided Crystal Ball (CB) functions~\cite{1981PhDT........26O}
with common mean $\mu_0$ and width $\sigma$, and different values of $n$ and $\alpha$.
The Crystal Ball function combines a Gaussian core and a power-law tail with an exponent $n$ to account for incomplete photon energy containment
in the cluster related to the material in front of the calorimeter, and for other reconstruction effects:
\ifthenelse{\boolean{cms@external}}{
\begin{multline}
f_\mathrm{CB}(\mu) = \Biggl\{ \\
\frac{N}{\sqrt{2\pi}\sigma}\exp\Biggl(-\frac{(\mu-\mu_0)^2}{2\sigma^2}\Biggr), \text{ for }  \frac{\mu-\mu_0}{\sigma}>\alpha; \\
\frac{N}{\sqrt{2\pi}\sigma}\Biggl( \frac{n}{\abs{\alpha}}\Biggr)^n \exp \Biggl(-\frac{\abs{\alpha}^2}{2}\Biggr) \left ( \frac{n}{\abs{\alpha}}- \abs{\alpha}-\frac{\mu-\mu_0}{\sigma}\right )^{-n},\\
\text{for }\frac{\mu-\mu_0}{\sigma} \le \alpha.
\label{cbfcn}
\end{multline}
}{
\begin{equation}
f_\mathrm{CB}(\mu) =  \begin{cases} \frac{N}{\sqrt{2\pi}\sigma}\exp\bigl(-\frac{(\mu-\mu_0)^2}{2\sigma^2}\bigr), &  \frac{\mu-\mu_0}{\sigma}>\alpha; \\ \frac{N}{\sqrt{2\pi}\sigma}\bigl( \frac{n}{\abs{\alpha}}\bigr)^n \exp \bigl(-\frac{\abs{\alpha}^2}{2}\bigr) \bigl( \frac{n}{\abs{\alpha}}- \abs{\alpha}-\frac{\mu-\mu_0}{\sigma}\bigr )^{-n},   &  \frac{\mu-\mu_0}{\sigma} \le \alpha.  \end{cases}
\label{cbfcn}
\end{equation}
}
The parameter $\alpha$ defines the transition between the Gaussian and power-law functions.
The fit to the response distribution in the first class of events for a simulated signal with $m_\PX = 150\GeV$ is shown in Fig.~\ref{fig:fitRes}.
The resolution in $\mu$ (\ie, the $\sigma$ in Eq.~(\ref{cbfcn})) improves by roughly 20\%, from 0.021 to 0.016, when the resonance mass increases from 150 to 850\GeV.
The resolution function $R$ is constructed so that $\sigma$ depends continuously on $m_\text{true}$ via a quadratic polynomial function.

\subsubsection{Parametric line shape}\label{sec:resp}
The theoretical line shape of unstable particles is modeled with a Breit--Wigner (BW)  distribution with a mass-dependent width~\cite{Heinemeyer:2013tqa}:
\begin{equation}
g_\text{BW}(m | m_\PX, \Gamma_\PX) = \frac{N}{(m^2-m_\PX^2)^2+m_\PX^2 \Gamma_\PX^2}.
 \label{eq:BW}
 \end{equation}
 The effect of the proton parton distribution functions (PDF) on the signal shape for a wide resonance
(with $\Gamma_\PX= 0.1 M_\PX$) is investigated with high-mass Higgs boson samples produced with the
\POWHEG generator~\cite{cps}. The shape obtained after convolution with the PDF is still well described
by a BW function within an accuracy better than 1\%.

\begin{figure}[t!bh]
 \begin{center}
\includegraphics[width=0.48\textwidth]{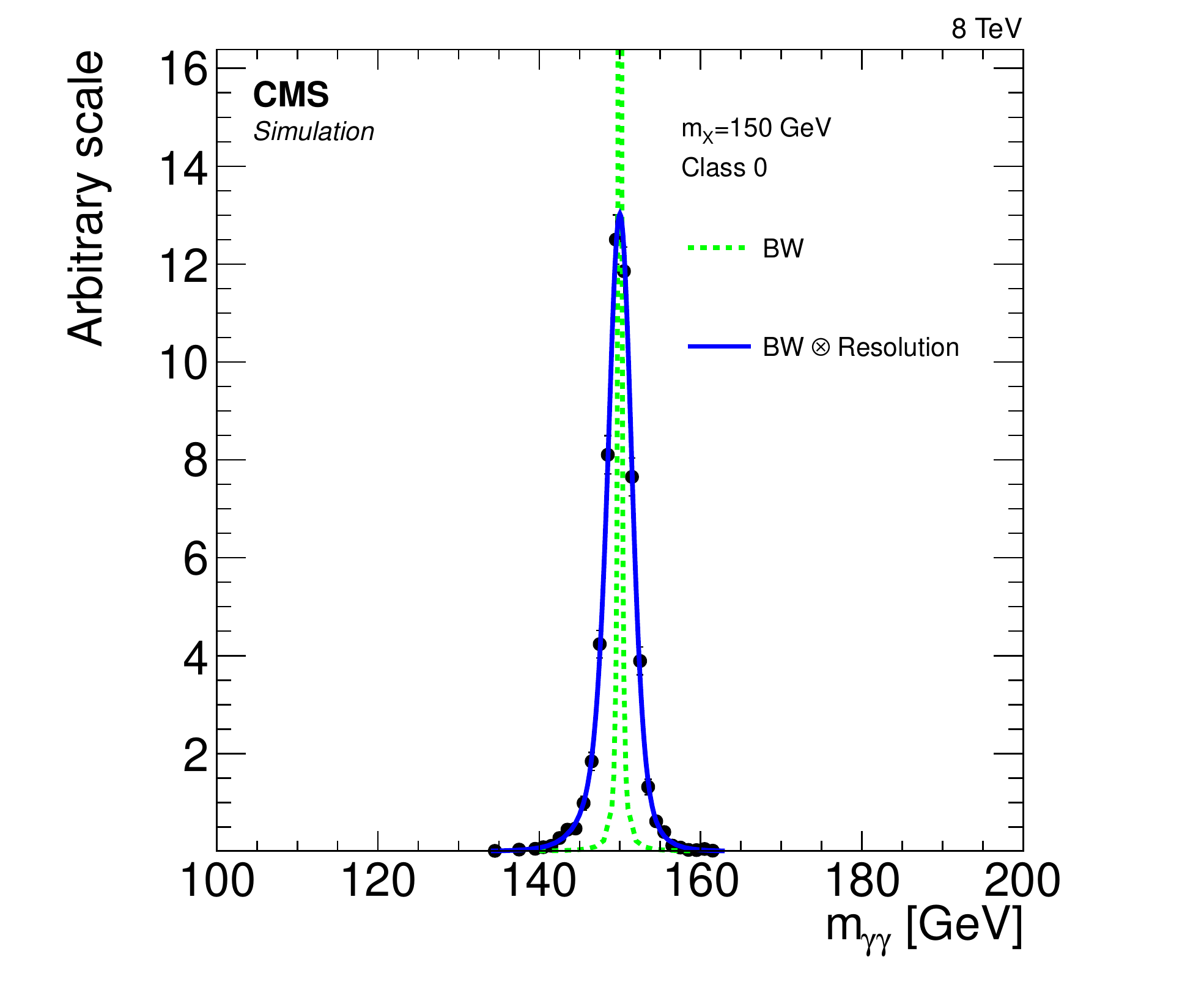}
\includegraphics[width=0.48\textwidth]{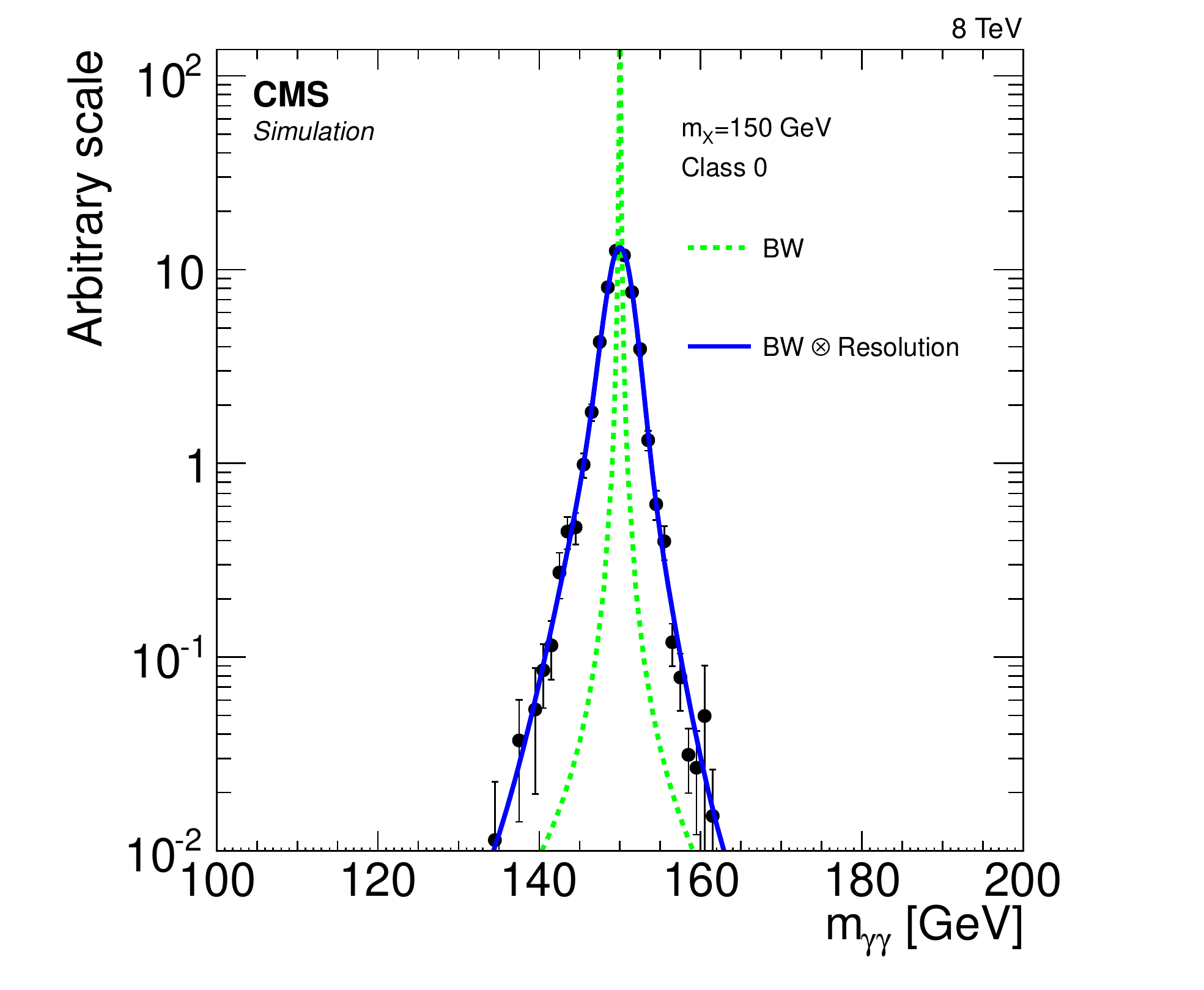}
  \caption{Parametrized signal shape for a signal with $m_\PX = 150\GeV$ and $\Gamma_\PX = 0.1\GeV$,
  displayed with linear (\cmsLeft) and logarithmic (\cmsRight) scales.
  The solid curve shows the result of a fit to the simulated events (points with error bars);
  the dashed curve represents the Breit--Wigner component of the model.}
 \label{fig:conv250}
 \end{center}
\end{figure}

The final signal model is obtained from the convolution of the BW in Eq.~(\ref{eq:BW}) and the response function $R$:
\begin{equation}
g_S(m_{\gamma \gamma}|m_\PX, \Gamma_\PX) = R(m_{\Pgg \Pgg}| m)\otimes g_\mathrm{BW} (m|m_\PX, \Gamma_\PX).
\label{eq:sigmod}
\end{equation}
This signal model is validated by fitting the reconstructed mass distribution for a simulated signal with $\Gamma_\PX = 0.1\GeV$, as shown in Fig.~\ref{fig:conv250}.

\subsection{Background modeling}\label{sec:bkg}
\begin{figure*}[tbh]
 \begin{center}
\includegraphics[width=0.48\textwidth]{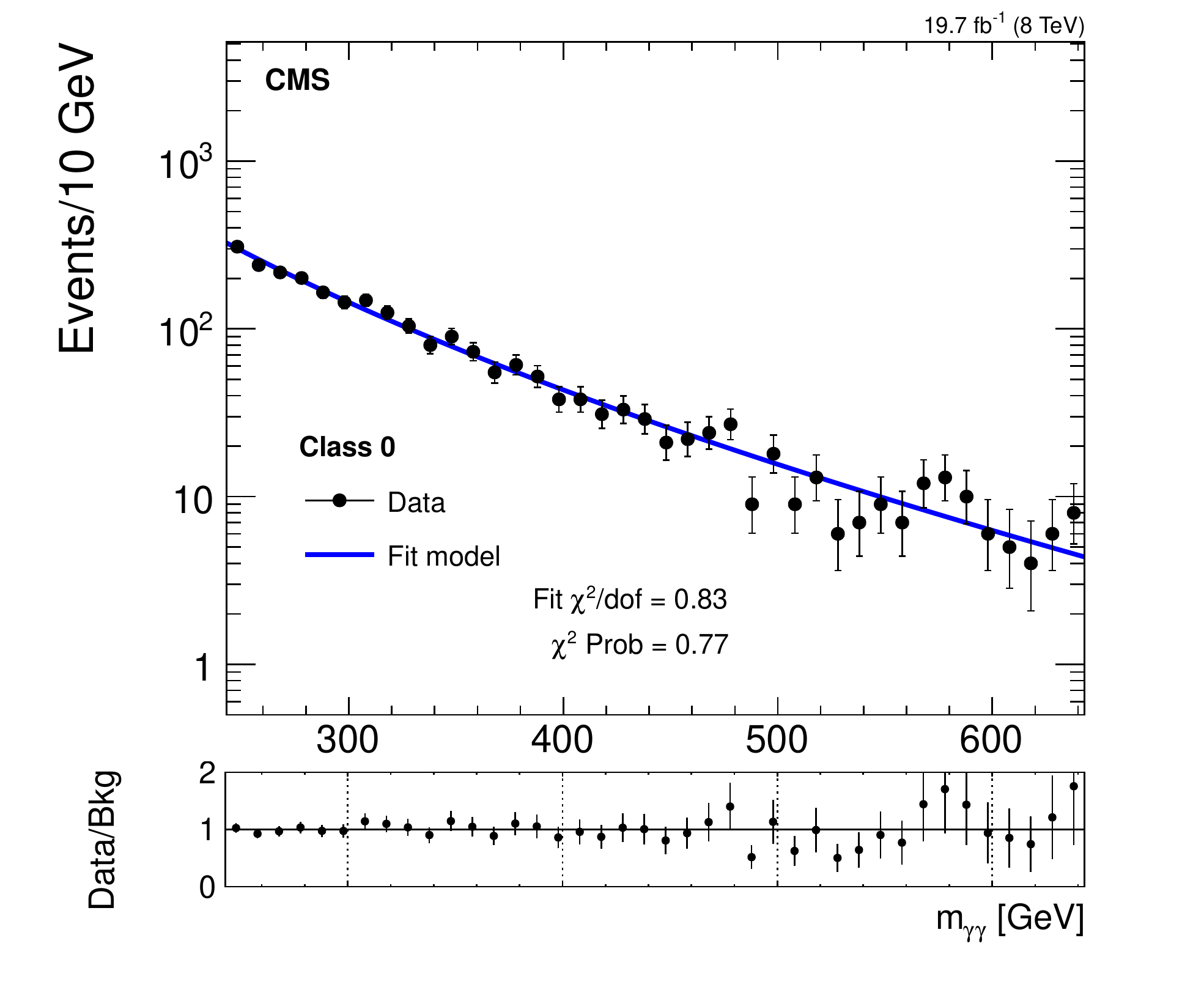}
\includegraphics[width=0.48\textwidth]{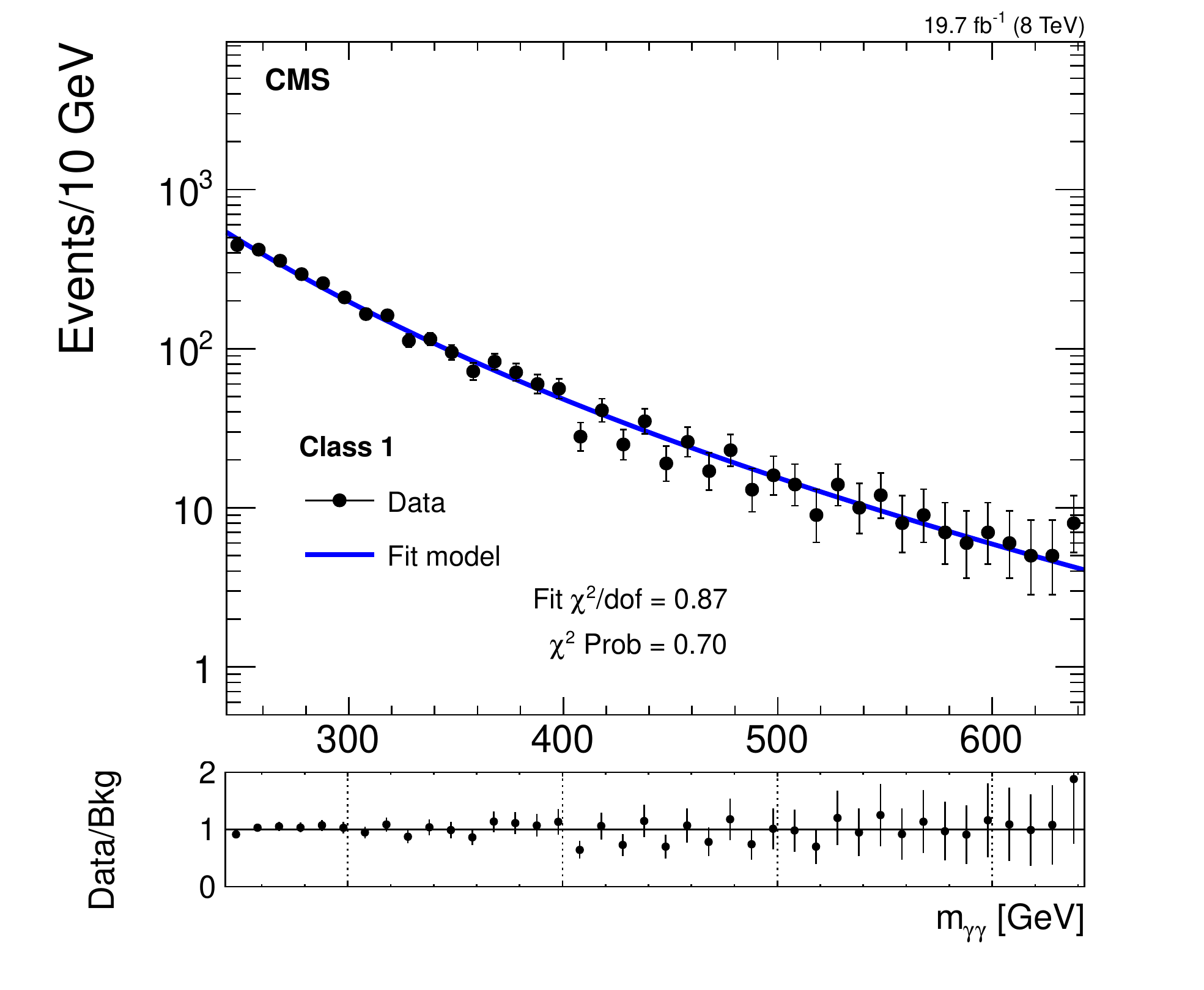}
\includegraphics[width=0.48\textwidth]{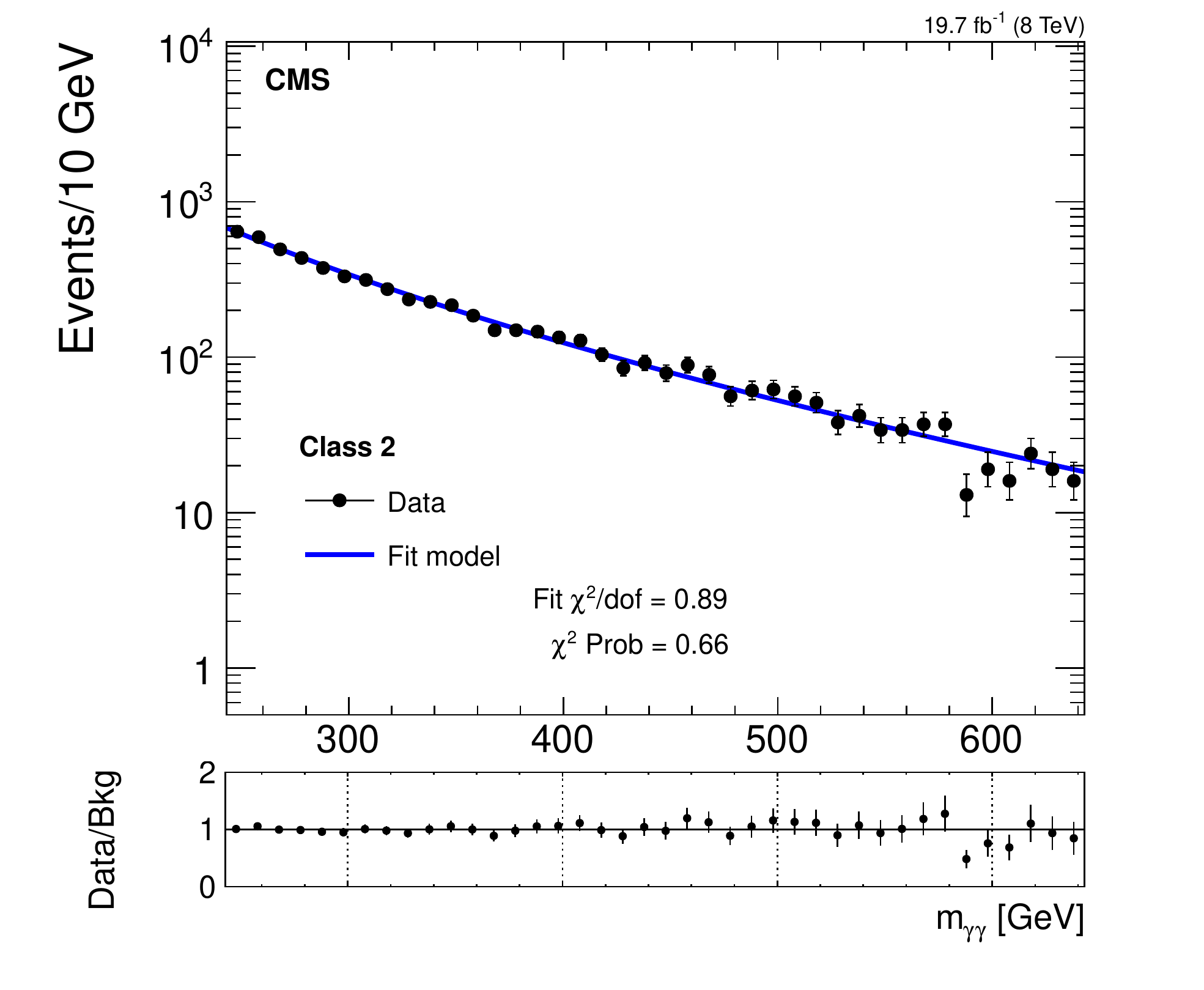}
\includegraphics[width=0.48\textwidth]{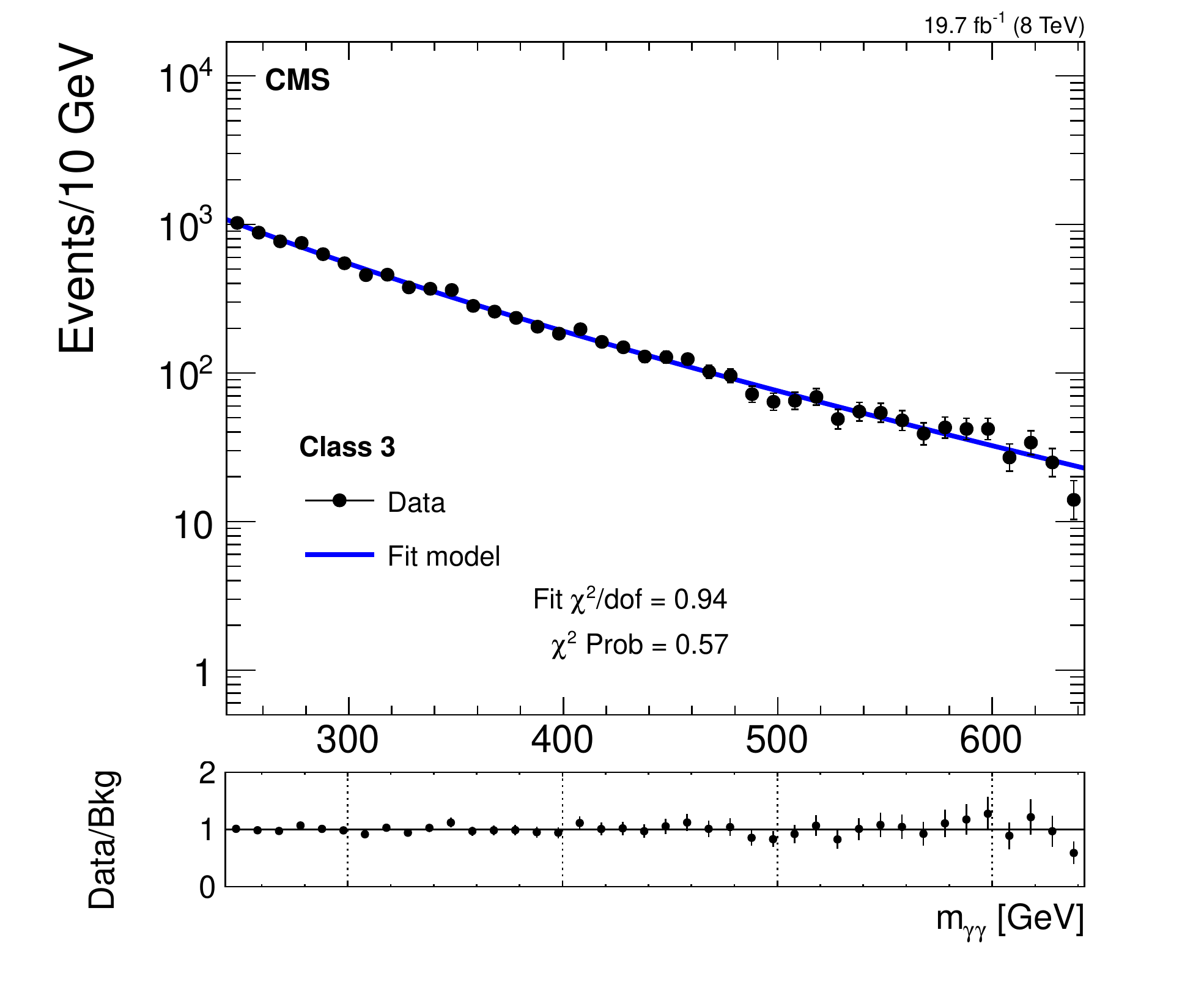}
  \caption{Fits to the diphoton mass distributions in the four classes of events in the window [240, 640] GeV chosen for searching for a peak near 350 GeV using the $f_0$ model and assuming no signal. The ratio between data and $f_0$ is displayed bin by bin in the bottom plot.}
 \label{fig:fitExpPAR}
 \end{center}
\end{figure*}

The modeling of the background relies entirely on the data, so there are no systematic uncertainties due to potential
mismodeling of the background processes by the MC simulation.
Because the exact functional form of the background in each event class is not known,
the parametric model must be flexible enough to describe a variety of potential underlying functions.
Using an incorrect background model can lead to biases in the measured signal yield, which can
strongly reduce the sensitivity of the analysis to any potential signal.
The procedure used to determine the background fitting function, which results in a negligible bias, is presented here.
In this study the bias on the fitted signal yield is defined as the difference between the number of fitted signal events and the number of expected signal events.
A set of analytical functions that could
describe the unknown true background distribution in data is first determined.
The five functions considered as possible truth models are analytical forms that are used in dijet resonance searches~\cite{dijet} to describe both data and QCD predictions and models that are frequently used in diphoton resonance searches~\cite{Khachatryan:2014ira}. The bias study procedure described below is performed considering each function among the available set of models. The candidate fitting function that performs best in this study is $f_0$, the product of exponential and power-law functions.
 \begin{equation}
  f_0(m) = \re^{-p_1 m}\, m^{-p_2} \label{eq:f0}.
 \end{equation}
As an example, the fits of this model to the data in the four event classes are shown in Fig. ~\ref{fig:fitExpPAR} for the fit range [240, 640] \GeV, used for
searching for a peak near 350 \GeV.
In order to check that a background fit model results in a negligible bias in the fitted signal yield, we construct pseudo-data sets
with the other four analytical models by randomly drawing diphoton mass values from them.
The number of background events generated in each set is equal to the number of events observed in data in a fixed mass range.
Background-only pseudo-data sets are then fitted with the signal + background probability density function.
The criterion for the bias to be negligible is that it must be five times smaller than the statistical uncertainty
 in the number of fitted signal events for all four types of pseudo-data sets across the entire mass region of interest.
When this criterion is satisfied, any potential bias from the background fit function
 can be neglected in comparison with the statistical uncertainty from the finite data sample.
The functional form of the truth model being fixed, the lower and upper bounds of the fit range are varied
 for each hypothetical resonance mass in order to minimize the bias.
The desired fit range for each hypothetical mass is the one that gives the smallest bias for all the truth models considered.
This study is performed for a resonance with width equal to 10\% of its mass, as the bias tends to increase with increasing resonance width.
The optimal lower and upper bounds of the fit range are parametrized as a function of the resonance mass with polynomial functions.

As a closure test, the obtained fit ranges are used to compute the bias on the number of fitted signal events as a function of the assumed resonance mass.
We find that the $f_0$ function produces a sufficiently small bias for all four truth
models in all event classes and for any width of the resonance up to 10\% of the mass.
We therefore use this background function to fit the data.

Our approach to extract the numbers of signal and background events cannot be used above $m_\PX = 850\GeV$
because of the very small number of events in the data.
This is therefore the highest value of mass considered in this search for new resonances.
The lowest value of $m_\PX$ considered is 150\GeV, and the fitted range in $m_{\gamma\gamma}$ is $[130, 1000]\GeV$.
The maximum value of the width of the resonance is fixed to 10\% of the resonance mass itself.
This value is limited by the width of the resonance mass points at the edges of the range ($150$ and $850$\GeV)
which have to be included in the fitting range ($[130, 1000]$\GeV) within at least one sigma.

\subsection{Systematic uncertainties}\label{sec:syst}
The experimental systematic uncertainties can be separated
into those related to the yield and those related to the signal shape.
A log-normal prior is assumed for the uncertainties in the class yields,  while
the shape uncertainties are incorporated as parametric variations of the model.

The normalization uncertainty related to the integrated luminosity is 2.6\%~\cite{CMS-PAS-LUM-13-001}.
The photon-related uncertainties are the same as in~\cite{Khachatryan:2014ira}.
The uncertainty in the energy scale is conservatively increased to 0.5\% in the barrel and to 2\% in the endcaps to take into account additional nonlinearities.
Systematic uncertainties related to individual photons are then propagated to the signal model, where they result in uncertainties in the peak position and width.
The 1\% (2.6\%) normalization uncertainty in the barrel (endcap) related to the offline photon identification
is taken from the largest uncertainty in the data/MC scale factors computed with $\Z\to\EE$ events using the tag-and-probe technique.
A 1\% normalization uncertainty is also assumed in the trigger efficiency.
A normalization uncertainty of 2--5\% is included to take into account
possible event class migration due to $\RNINE$ selection efficiency
uncertainties in both barrel and endcap.
More details about trigger and $\RNINE$ efficiencies can be found
in Ref.~\cite{Khachatryan:2014ira}.

Systematic uncertainties estimated for the SM-like Higgs boson
search~\cite{Khachatryan:2014ira} can
be safely used for this analysis as well for low mass resonances,
where the bulk of the photon $\pt$ distribution is close to that of
the photons coming from decays of the SM-like Higgs boson.
This is not the case for high-mass resonances.
Therefore a normalization uncertainty of 5\% per photon pair is also
included to account for the differences in the $\pt$ spectra
of the signal photons and the electrons from $\Z\to\EE$ used to
estimate the uncertainties~\cite{Chatrchyan:2012oaa}.

The use of the BW model has an accuracy of the order $\Gamma_\PX/m_\PX$~\cite{Heinemeyer:2013tqa}.
A global normalization uncertainty equal to $\Gamma_\PX/m_\PX$ is added
to account for any uncertainty due to the theoretical signal line shape.
Table~\ref{Tab:systematics baseline} lists all the systematic uncertainties accounted for in the analysis.

\begin{table*}[bthp]
\topcaption{Summary of uncertainties that have impact on the signal strength, applicable to events in all classes.
}
\begin{center}
\begin{tabular}{c|cc}
{{Sources of systematic uncertainty}} & \multicolumn{2}{ c }{{Uncertainty}}\\
\hline
\hline
{{Per photon} } & {Barrel} & {Endcap} \\
\hline
Energy resolution,  $\RNINE > 0.94$ (low $\eta$, high $\eta$) & \small{0.10\%, 0.20\%} & \small{0.14\%, 0.06\%} \\
Energy resolution,  $\RNINE < 0.94$ (low $\eta$, high $\eta$) & \small{0.10\%, 0.18\%} & \small{0.18\%, 0.12\%} \\
\hline
Photon energy scale &  0.5\% & 2\% \\
\hline
{Photon identification efficiency} & 1.0\% & 2.6\%\\
\hline
\hline
{{Per event}} & {Barrel} & {Endcap}\\
\hline
{Integrated luminosity} & 2.6\% & 2.6\% \\
{Vertex finding efficiency} & 0.2\% & 0.2\% \\
{Trigger efficiency} & 1.0\% & 1.0\%\\
{$\RNINE$ class migration} & 2.3\% & 5.5\%\\
\hline
{Additional normalization uncertainty} & 5\% & 5\%  \\
\hline
{Breit--Wigner model} & 0.01--10\% & 0.01--10\% \\
\end{tabular}
\end{center}
\label{Tab:systematics baseline}
\end{table*}

\section{Results and interpretation}\label{sec:res}
\begin{figure}[tbh]
 \begin{center}
\includegraphics[width=0.48\textwidth]{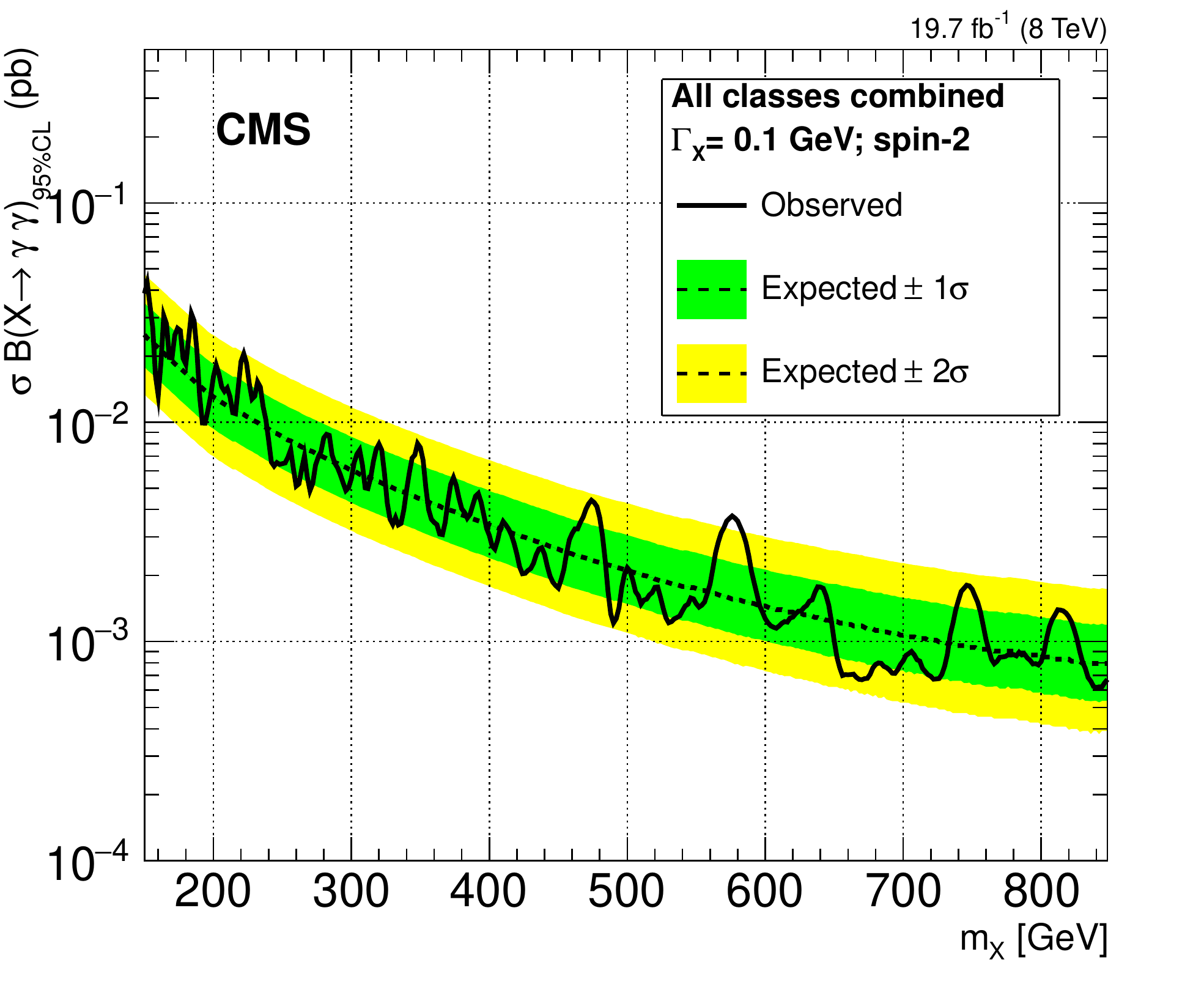}
  \caption{Exclusion limit at 95\% \CL on the cross section times branching fraction of a new, narrow, spin-2 resonance decaying into two photons
 as a function of the resonance mass hypothesis, combining the four classes of events. }
 \label{fig:combinedGRAVITON}
 \end{center}
\end{figure}

\begin{figure}[tbh]
 \begin{center}
\includegraphics[width=0.48\textwidth]{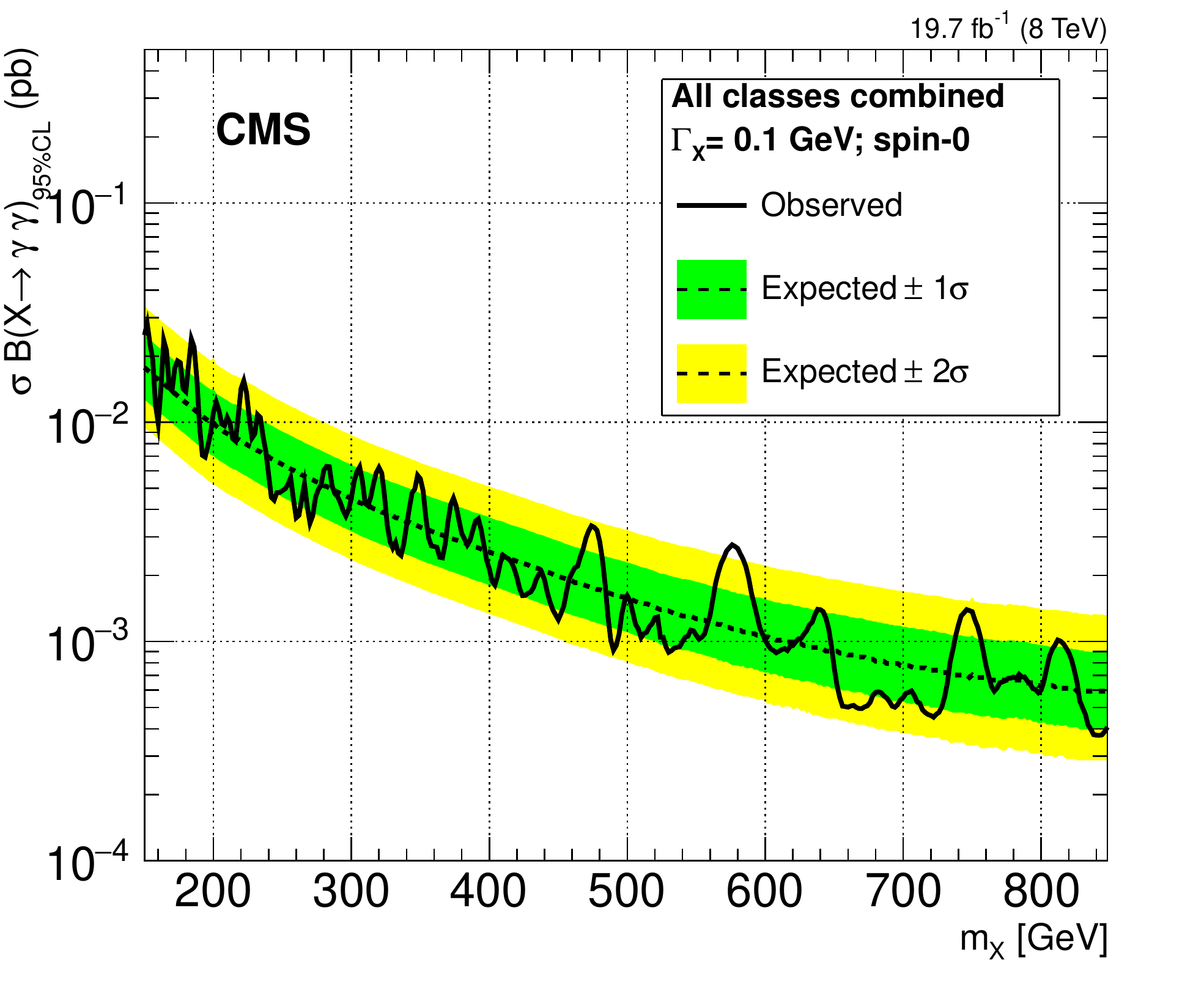}
\includegraphics[width=0.48\textwidth]{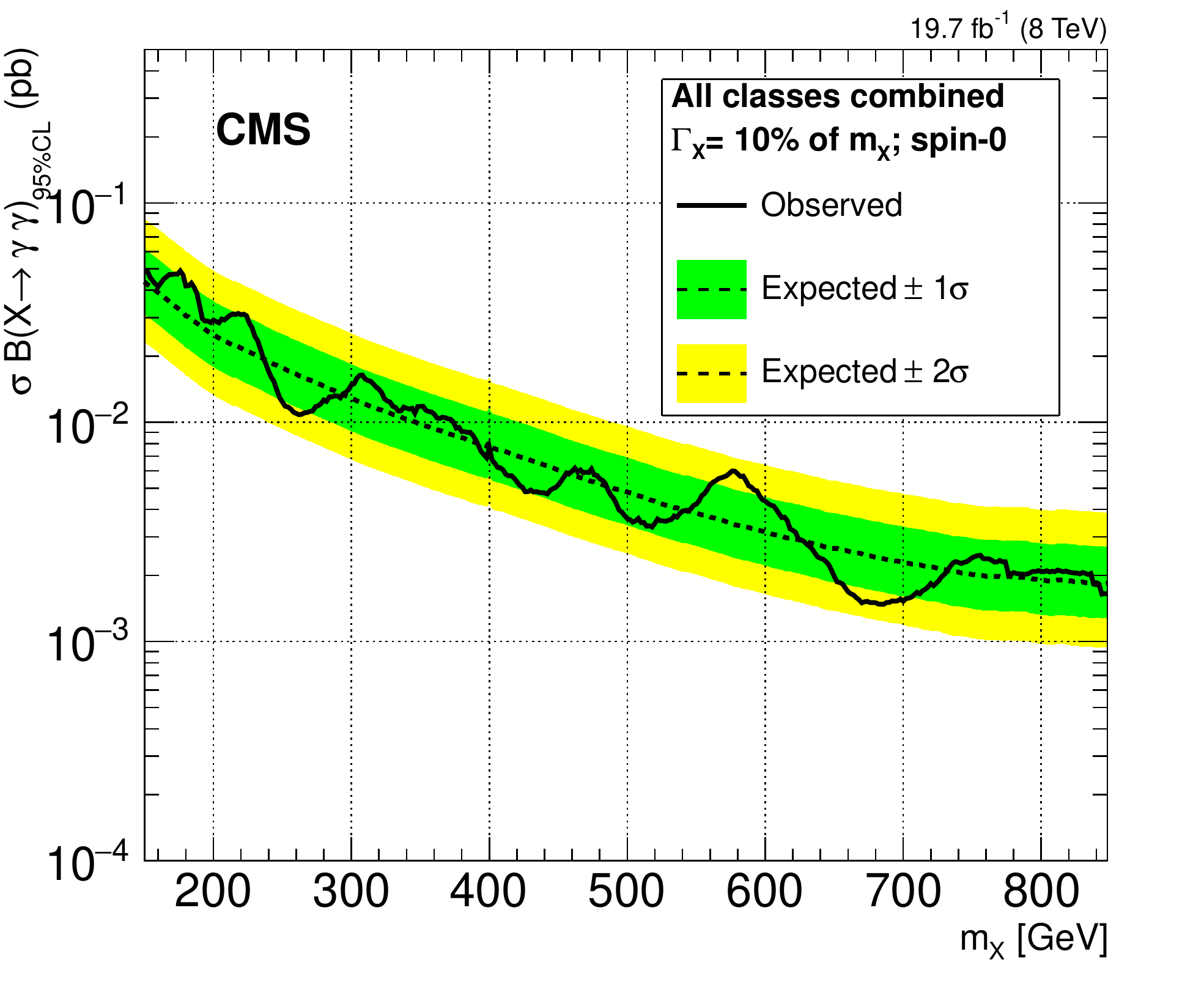}
  \caption{Exclusion limit at 95\%\ \CL on the cross section times branching fraction of a new, spin-0 resonance decaying into two photons
 as a function of the resonance mass hypothesis, combining the four classes of events. The results for a narrow resonance hypothesis ($\Gamma_\PX = 0.1\GeV$) (\cmsLeft) and for a wide resonance hypothesis ($\Gamma_\PX = 0.1 m_\PX$) (\cmsRight) are shown.}
 \label{fig:combined}
 \end{center}
\end{figure}

\begin{figure}[tbh]
 \begin{center}
\includegraphics[width=0.48\textwidth]{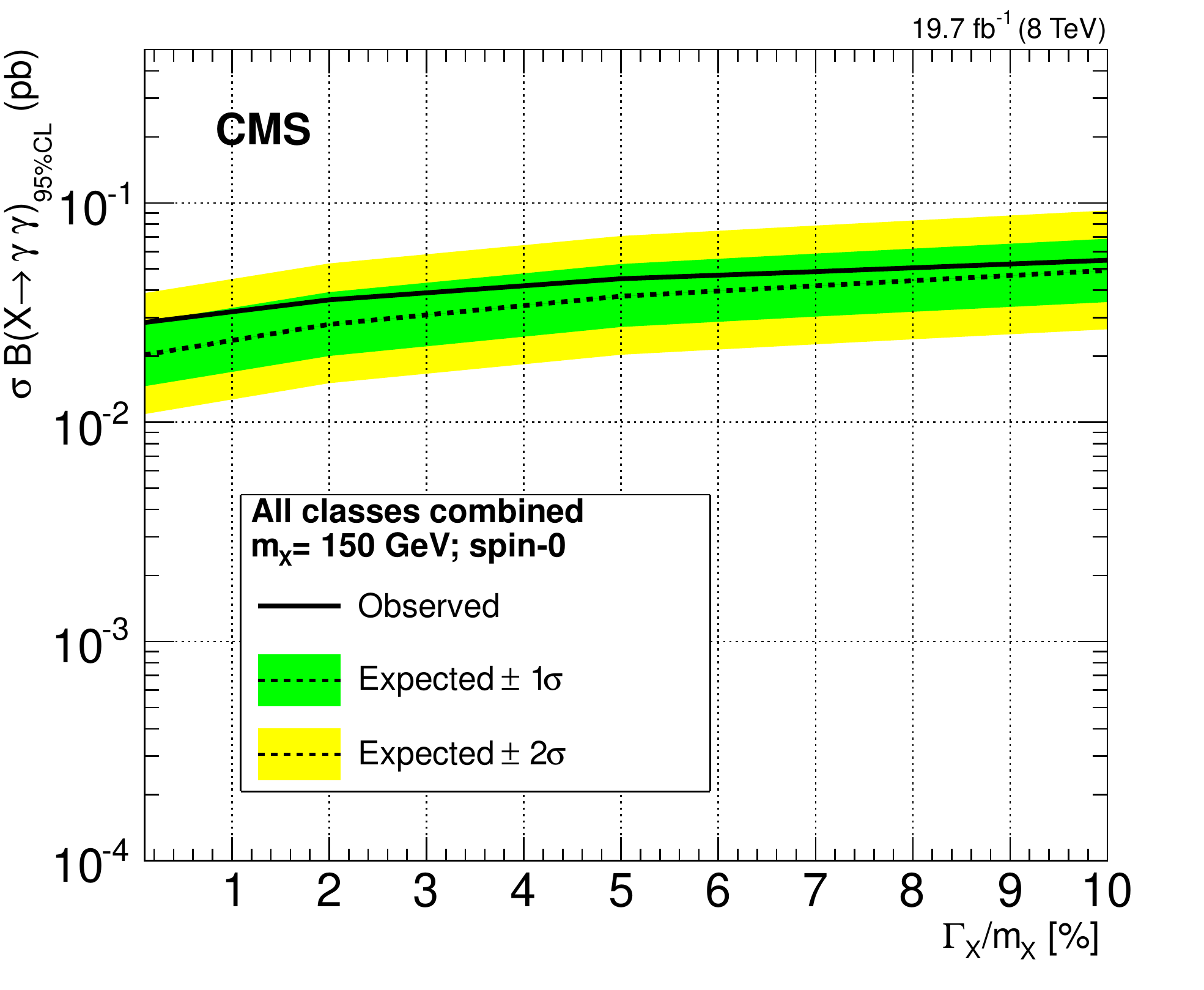}
\includegraphics[width=0.48\textwidth]{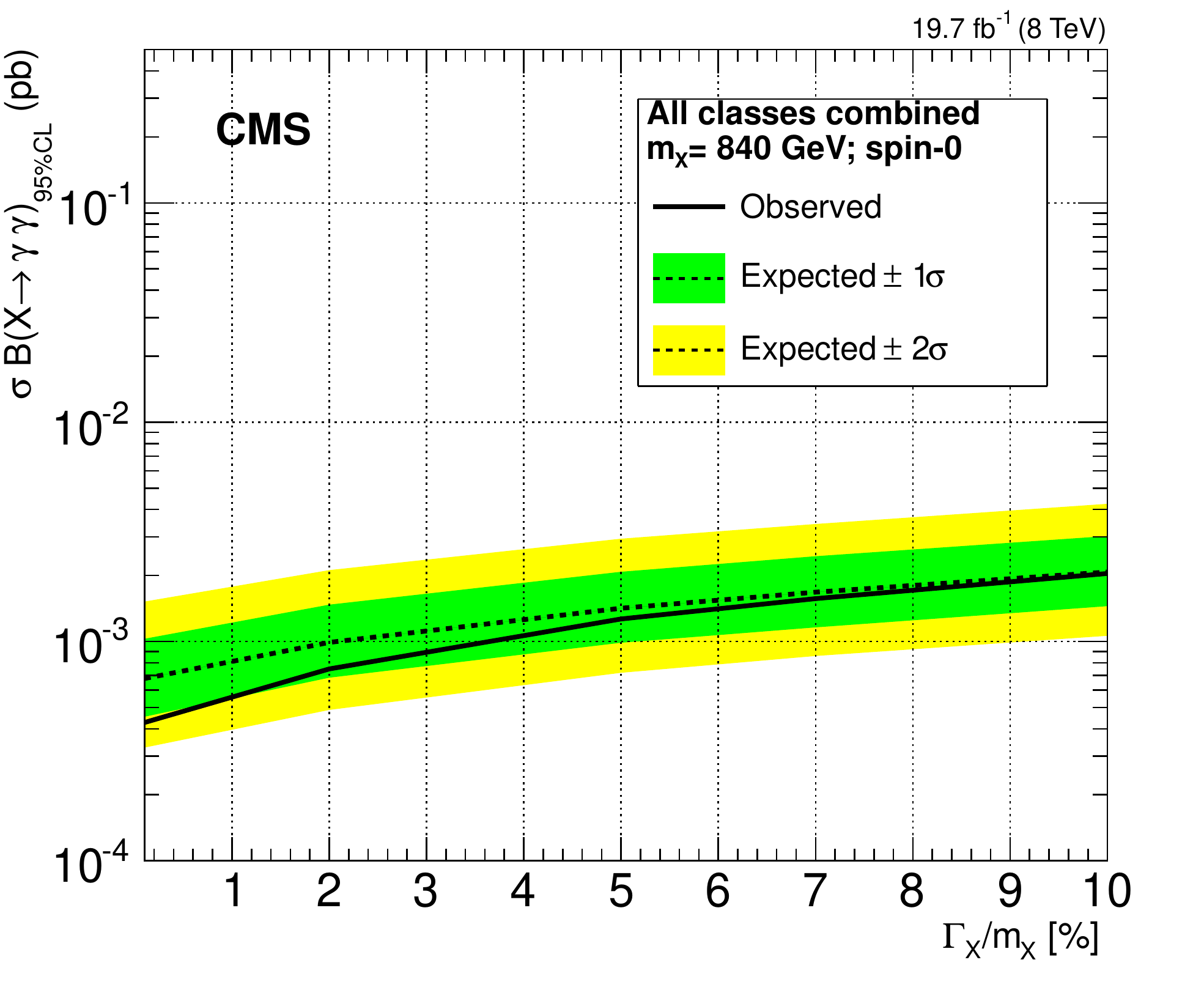}
  \caption{Exclusion limit at 95\% \CL on the cross section times branching fraction of a new, spin-0 resonance decaying into two photons
 as a function of the resonance width hypothesis, combining the four classes of events. The results for  $m_\PX $ = 150 (840) \GeV are shown  \cmsLeft (\cmsRight).}
 \label{fig:combinedVsWidth}
 \end{center}
\end{figure}

\begin{figure}[tbh]
 \begin{center}
\includegraphics[width=0.48\textwidth]{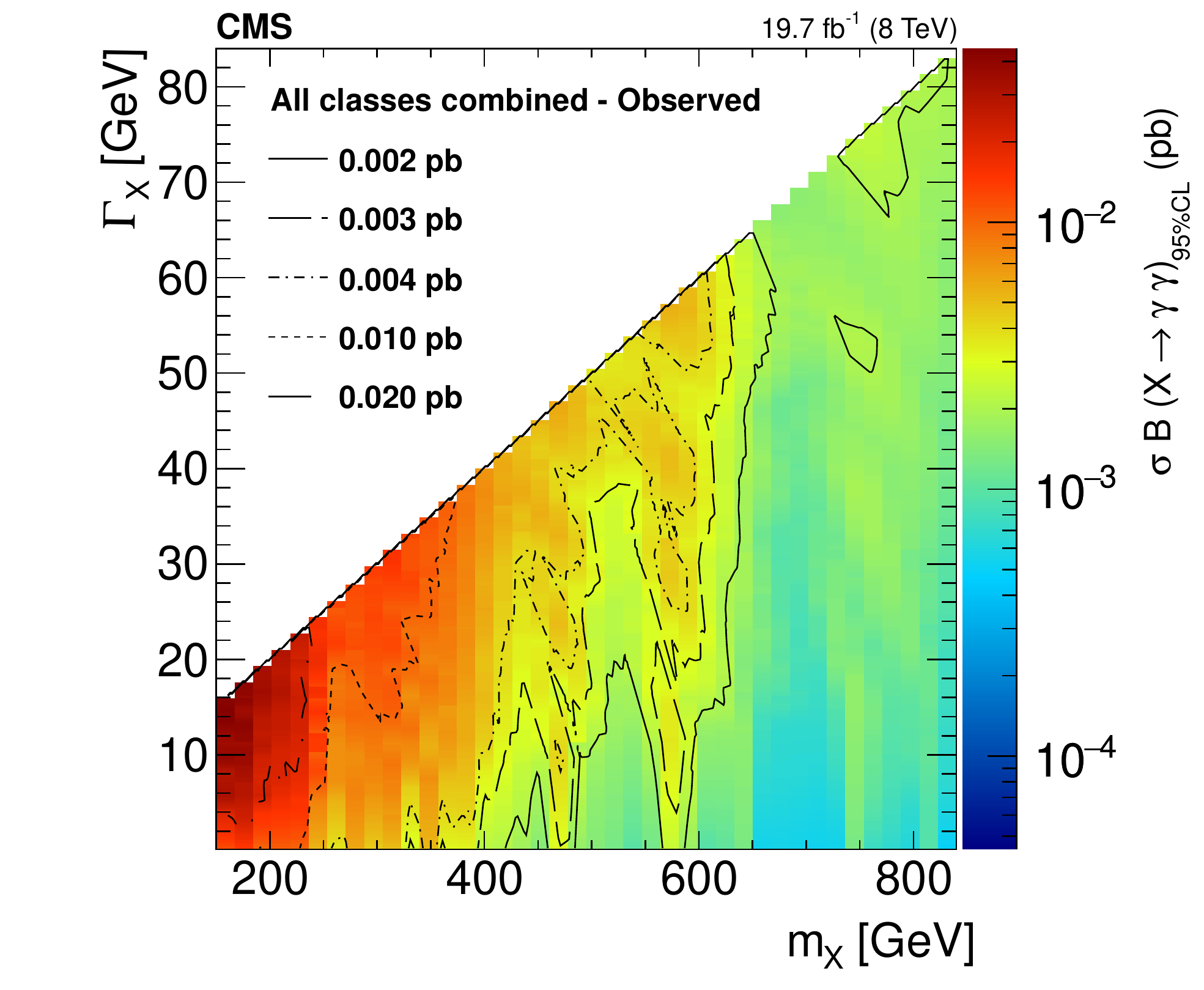}
\includegraphics[width=0.48\textwidth]{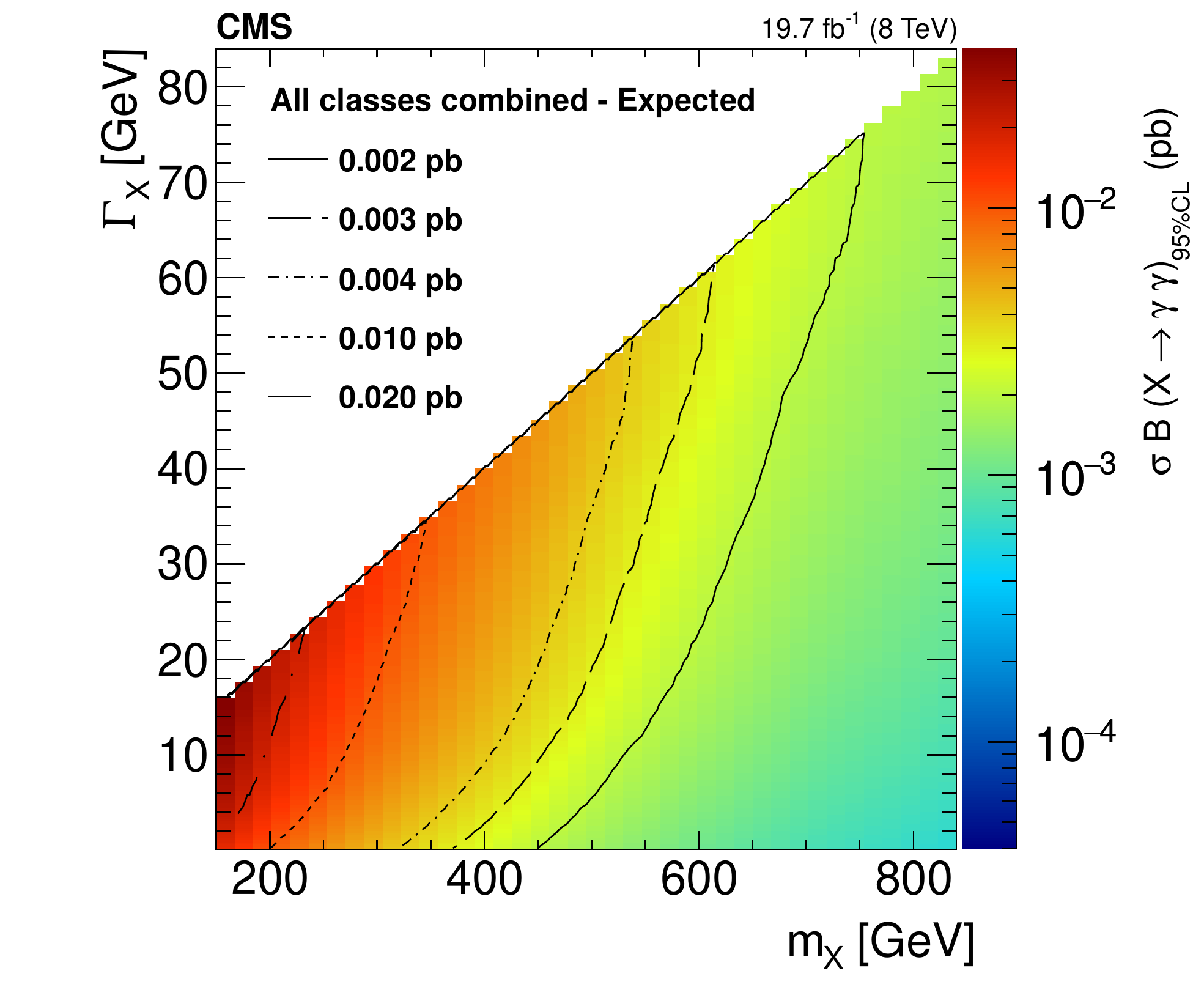}
  \caption{ Combined observed (\cmsLeft) and expected (\cmsRight) exclusion limits at 95\% \CL on the cross section times branching fraction of a new, spin-0 resonance decaying into two photons
 as a function of the resonance mass and width hypotheses. Contours are displayed for different values of $\sigma\mathcal{B}(X\to
  \gamma \gamma)$.}
 \label{fig:2Dcomb}
 \end{center}
\end{figure}

The invariant mass spectra show no clear evidence for the presence of a new particle decaying to two photons.
Exclusion limits are therefore computed. A modified frequentist \CLs method~\cite{Read:2000ru} is employed,
with the asymptotic approximation for the statistic test as described in Ref.~\cite{Cowan:2010js}.
Model-independent results are presented for a spin-0 and spin-2 resonance produced via gluon-gluon fusion.
Figure~\ref{fig:combinedGRAVITON} shows the 95\% confidence level (\CL) exclusion limits on the production cross section times branching fraction ($\sigma \mathcal{B}$),
obtained by combining all four event classes, as a function of mass for a narrow ($\Gamma_\PX = 0.1\GeV$) spin-2 resonance.
Figure~\ref{fig:combined} shows the 95\% \CL combined limits for two width hypotheses,
$\Gamma_\PX = 0.1\GeV$ and $\Gamma_\PX = 0.1 m_\PX$, as a function of mass for the spin-0 model.
The results shown in Figs.~\ref{fig:combinedGRAVITON} and \ref{fig:combined}~(\cmsLeft) are
similar because of the small differences
in the efficiency times the acceptance in the two different spin scenarios.
Figure~\ref{fig:combinedVsWidth} shows the dependence of the limit on $\Gamma_\PX$
for two values of the resonance mass (150 and 840\GeV) in the spin-0 model.
Figure~\ref{fig:2Dcomb} shows the 95\% \CL exclusion limits on the production
cross section times branching fraction as a function of the resonance mass and
width in the spin-0 model. The expected limits lie between $6 \times 10^{-4} \unit{pb}$  and $4\times 10^{-2} \unit{pb}$
 over the full mass range analyzed.
The observed limits are consistent with the expected sensitivity of the analysis in the no signal hypothesis.
The largest excess  is observed at $m_\PX \sim 580\GeV$ with a local significance of less than $2.5\sigma$.

\subsection{Interpretation in two-Higgs-doublet model}\label{sec:2HDM}
In this section the model-independent limits obtained for a hypothetical heavy diphoton resonance are interpreted
 in the context of the two heavy neutral Higgs bosons, \PH and \PA, predicted in the 2HDM~\cite{Dev:2014yca}.
In this model, the production cross sections for \PH and \PA, as well as the branching
fractions for their decays to two photons depend on two parameters, $\alpha$ and $\beta$. The mixing angle between
\PH and \Ph is given by $\alpha$, while $\tan\beta$ is the ratio of the vacuum expectation values of the two Higgs doublets.
The 2HDM cross sections are calculated by means of the \textsc{SusHi}~\cite{Harlander:2012pb} program, and branching fractions are obtained using 2HDMC~\cite{Eriksson:2009ws}.

Exclusion regions in the $\tan\beta$ versus $\cos(\beta-\alpha)$ plane are shown only for the diphoton decay of the pseudoscalar Higgs boson \PA;
no region of the phase space can be excluded for the decay of the heavy H scalar.

\begin{figure}[tbh]
 \begin{center}
\includegraphics[width=0.48\textwidth]{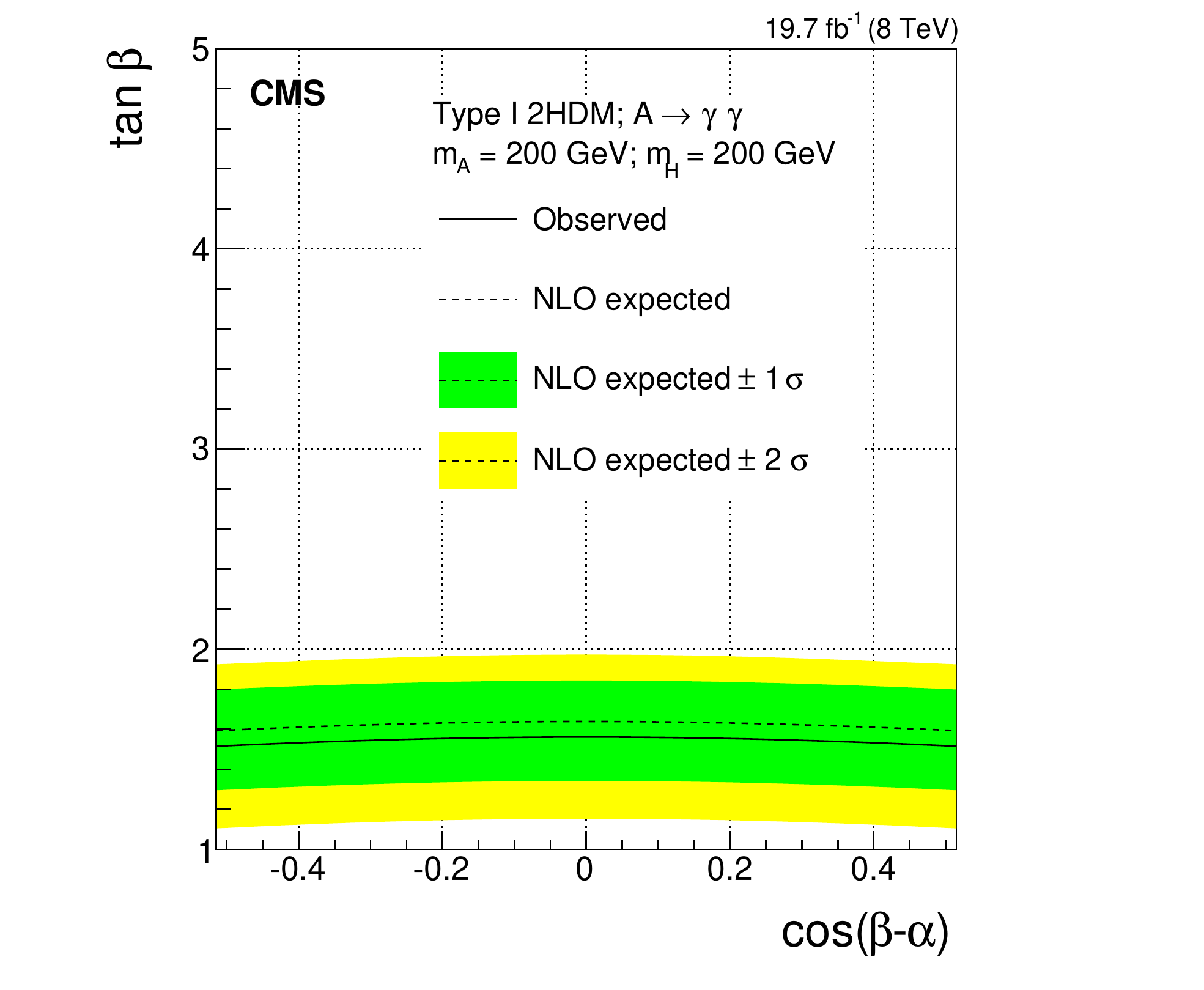}
\includegraphics[width=0.48\textwidth]{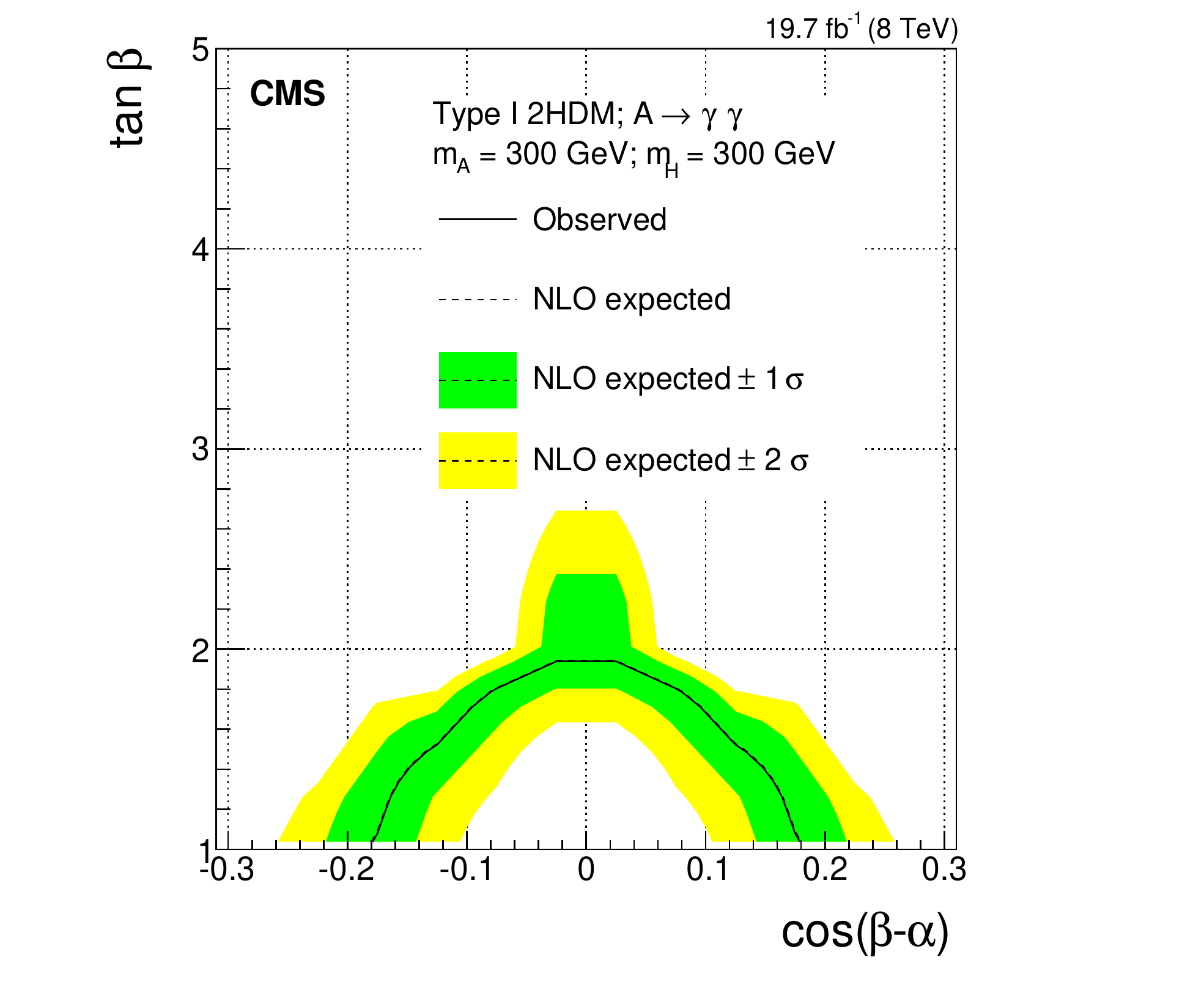}
   \caption{Observed and expected 95\% \CL exclusion regions for gluon-fusion production of a
   heavy Higgs boson \PA of mass 200\GeV (\cmsLeft) and 300\GeV (\cmsRight) in the $\tan\beta$ versus
    $\cos(\beta-\alpha)$ plane for the Type I 2HDM, assuming the \PH boson to be degenerate in mass with \PA. The regions below the curves are excluded. }
 \label{fig:mHbig}
 \end{center}
\end{figure}

Figure~\ref{fig:mHbig} shows the observed and expected exclusion regions for a heavy Higgs boson \PA of mass 200 and 300\GeV for the Type I 2HDM.
The case where \PH and \PA are degenerate in mass is considered.

In Fig.~\ref{fig:mHbig} the region below the curve is excluded.
These contour plots are similar to those in~\cite{Craig:2013hca}.
The constraints obtained in this analysis on the 2HDM parameters are complementary to those already set by ATLAS on heavy neutral Higgs boson production using multilepton final states~\cite{tagkey2015163} and by CMS on heavy neutral Higgs bosons production using multilepton and diphoton final states~\cite{Khachatryan:2015lba,Khachatryan:2014jya}.

\section{Summary}
A search for resonant production of two photons is performed using 19.7\fbinv of pp collisions collected
at $\sqrt{s}=8\TeV$, in the mass range 150--850\GeV.
Widths of the resonance $\PX$ in the range 0.1\GeV to $0.1 m_\PX$ are investigated.
Both spin-0 and spin-2 scenarios are considered.
A fit to the diphoton invariant mass distribution in data is performed using a parametric model for the signal and a background shape obtained directly from data.
No evidence for a signal is observed, and upper exclusion limits at 95\% \CL are set on the production cross section times branching fraction.
The model-independent upper limits extend over considerably wider mass and width ranges than in previous searches.
We further interpret these limits in the context of the 2HDM, presenting exclusion contours in the \tanb  versus \cosba plane.
This is the first search for heavy diphoton resonances carried out at the LHC to be interpreted in terms of the 2HDM.
\clearpage
\begin{acknowledgments}
\hyphenation{Bundes-ministerium Forschungs-gemeinschaft Forschungs-zentren} We congratulate our colleagues in the CERN accelerator departments for the excellent performance of the LHC and thank the technical and administrative staffs at CERN and at other CMS institutes for their contributions to the success of the CMS effort. In addition, we gratefully acknowledge the computing centers and personnel of the Worldwide LHC Computing Grid for delivering so effectively the computing infrastructure essential to our analyses. Finally, we acknowledge the enduring support for the construction and operation of the LHC and the CMS detector provided by the following funding agencies: the Austrian Federal Ministry of Science, Research and Economy and the Austrian Science Fund; the Belgian Fonds de la Recherche Scientifique, and Fonds voor Wetenschappelijk Onderzoek; the Brazilian Funding Agencies (CNPq, CAPES, FAPERJ, and FAPESP); the Bulgarian Ministry of Education and Science; CERN; the Chinese Academy of Sciences, Ministry of Science and Technology, and National Natural Science Foundation of China; the Colombian Funding Agency (COLCIENCIAS); the Croatian Ministry of Science, Education and Sport, and the Croatian Science Foundation; the Research Promotion Foundation, Cyprus; the Ministry of Education and Research, Estonian Research Council via IUT23-4 and IUT23-6 and European Regional Development Fund, Estonia; the Academy of Finland, Finnish Ministry of Education and Culture, and Helsinki Institute of Physics; the Institut National de Physique Nucl\'eaire et de Physique des Particules~/~CNRS, and Commissariat \`a l'\'Energie Atomique et aux \'Energies Alternatives~/~CEA, France; the Bundesministerium f\"ur Bildung und Forschung, Deutsche Forschungsgemeinschaft, and Helmholtz-Gemeinschaft Deutscher Forschungszentren, Germany; the General Secretariat for Research and Technology, Greece; the National Scientific Research Foundation, and National Innovation Office, Hungary; the Department of Atomic Energy and the Department of Science and Technology, India; the Institute for Studies in Theoretical Physics and Mathematics, Iran; the Science Foundation, Ireland; the Istituto Nazionale di Fisica Nucleare, Italy; the Ministry of Science, ICT and Future Planning, and National Research Foundation (NRF), Republic of Korea; the Lithuanian Academy of Sciences; the Ministry of Education, and University of Malaya (Malaysia); the Mexican Funding Agencies (CINVESTAV, CONACYT, SEP, and UASLP-FAI); the Ministry of Business, Innovation and Employment, New Zealand; the Pakistan Atomic Energy Commission; the Ministry of Science and Higher Education and the National Science Centre, Poland; the Funda\c{c}\~ao para a Ci\^encia e a Tecnologia, Portugal; JINR, Dubna; the Ministry of Education and Science of the Russian Federation, the Federal Agency of Atomic Energy of the Russian Federation, Russian Academy of Sciences, and the Russian Foundation for Basic Research; the Ministry of Education, Science and Technological Development of Serbia; the Secretar\'{\i}a de Estado de Investigaci\'on, Desarrollo e Innovaci\'on and Programa Consolider-Ingenio 2010, Spain; the Swiss Funding Agencies (ETH Board, ETH Zurich, PSI, SNF, UniZH, Canton Zurich, and SER); the Ministry of Science and Technology, Taipei; the Thailand Center of Excellence in Physics, the Institute for the Promotion of Teaching Science and Technology of Thailand, Special Task Force for Activating Research and the National Science and Technology Development Agency of Thailand; the Scientific and Technical Research Council of Turkey, and Turkish Atomic Energy Authority; the National Academy of Sciences of Ukraine, and State Fund for Fundamental Researches, Ukraine; the Science and Technology Facilities Council, UK; the US Department of Energy, and the US National Science Foundation.

Individuals have received support from the Marie-Curie program and the European Research Council and EPLANET (European Union); the Leventis Foundation; the A. P. Sloan Foundation; the Alexander von Humboldt Foundation; the Belgian Federal Science Policy Office; the Fonds pour la Formation \`a la Recherche dans l'Industrie et dans l'Agriculture (FRIA-Belgium); the Agentschap voor Innovatie door Wetenschap en Technologie (IWT-Belgium); the Ministry of Education, Youth and Sports (MEYS) of the Czech Republic; the Council of Science and Industrial Research, India; the HOMING PLUS program of the Foundation for Polish Science, cofinanced from European Union, Regional Development Fund; the Compagnia di San Paolo (Torino); the Consorzio per la Fisica (Trieste); MIUR project 20108T4XTM (Italy); the Thalis and Aristeia programs cofinanced by EU-ESF and the Greek NSRF; and the National Priorities Research Program by Qatar National Research Fund.
\end{acknowledgments}
\bibliography{auto_generated}
\cleardoublepage \appendix\section{The CMS Collaboration \label{app:collab}}\begin{sloppypar}\hyphenpenalty=5000\widowpenalty=500\clubpenalty=5000\textbf{Yerevan Physics Institute,  Yerevan,  Armenia}\\*[0pt]
V.~Khachatryan, A.M.~Sirunyan, A.~Tumasyan
\vskip\cmsinstskip
\textbf{Institut f\"{u}r Hochenergiephysik der OeAW,  Wien,  Austria}\\*[0pt]
W.~Adam, E.~Asilar, T.~Bergauer, J.~Brandstetter, E.~Brondolin, M.~Dragicevic, J.~Er\"{o}, M.~Flechl, M.~Friedl, R.~Fr\"{u}hwirth\cmsAuthorMark{1}, V.M.~Ghete, C.~Hartl, N.~H\"{o}rmann, J.~Hrubec, M.~Jeitler\cmsAuthorMark{1}, V.~Kn\"{u}nz, A.~K\"{o}nig, M.~Krammer\cmsAuthorMark{1}, I.~Kr\"{a}tschmer, D.~Liko, T.~Matsushita, I.~Mikulec, D.~Rabady\cmsAuthorMark{2}, B.~Rahbaran, H.~Rohringer, J.~Schieck\cmsAuthorMark{1}, R.~Sch\"{o}fbeck, J.~Strauss, W.~Treberer-Treberspurg, W.~Waltenberger, C.-E.~Wulz\cmsAuthorMark{1}
% ...
\vskip\cmsinstskip
% ...
1:~~Also at Vienna University of Technology, Vienna, Austria\\
2:~~Also at CERN, European Organization for Nuclear Research, Geneva, Switzerland\\
% ...

\end{sloppypar}
\end{document}